\documentstyle[12pt]{article}

\begin{document}

\setcounter{page}{1}

\pagestyle{plain} \vspace{1cm}
\begin{center}
\Large{\bf A Non-minimally Coupled Quintom Dark Energy Model on the Warped DGP Brane}\\
\small \vspace{1cm} {\bf Kourosh
Nozari$^{a,}$\footnote{knozari@umz.ac.ir}}, \quad {\bf M. R.
Setare$^{b,}$\footnote{rezakord@ipm.ir}},\quad
{\bf Tahereh Azizi$^{a,}$\footnote{t.azizi@umz.ac.ir}}\quad and \quad {\bf Noushin Behrouz$^{c}$} \\
\vspace{0.5cm} $^{a}${\it Department of Physics, Faculty of Basic
Sciences,\\
University of Mazandaran,\\
P. O. Box 47416-95447, Babolsar, IRAN}\\
$^{b}$ {\it Department of Science, Payam-e Nour University, Bijar, IRAN}\\
{\it $^{c}$ Department of Physics, Payam-e Nour University,\\
P. O. Box 919, Mashad, IRAN}

\end{center} \vspace{1.5cm}
\begin{abstract}
We study dynamics of equation of state parameter for a non-minimally
coupled quintom dark energy component on the warped DGP brane. We
investigate crossing of the cosmological constant line in this
scenario. This crossing occurs in both DGP$^{\pm}$ branches of the
model.
\end{abstract}
\vspace{2cm}
\newpage
\section{Introduction}
In the last few years, an increasing number of astronomical
observations (such as data from CMB temperature fluctuations
spectrum and Supernova type Ia redshift-distance surveys) have
indicated that current universe is almost flat and undergoing a
positively accelerated phase of expansion [1]. This phenomenon is
not predicted by standard cosmology governed by general relativity
with the known matter constituents. To explain this cosmic positive
acceleration, mysterious dark energy has been proposed. There are
several dark energy models which can be distinguished by, for
instance, their equation of state (EoS)
$(\omega_{de}=\frac{P_{de}}{\rho_{de}}) $ during the evolution of
the universe. The cosmological constant is the simplest model of
dark energy with equation of state $ \omega=-1$, but the huge
fine-tuning required for its magnitude makes it unfavorable for the
cosmologists. It is then tempting to find alternative models of dark
energy which have dynamical nature unlike cosmological constant. An
accelerated expansion can be realized by using a scalar field whose
origin may be found in superstring or supergravity theories; some of
these models are quintessence, k-essence, tachyonic models,
dilatonic models and phantom fields [2]( see also [3]). Other
alternative approaches to accommodate dark energy are modification
of general relativity by considering additional spatial dimensions
[4-6] or modified Einstein-Hilbert action [7].

In recent years, there has been a lot of interest in the extra
dimensional theories by modifying the old Kaluza-Klien picture,
where the extra dimensions must be sufficiently compact. These
recent developments are based on the idea that ordinary matter and
gauge fields could be confined to a three dimensional world (brane),
while gravity and possibly non-standard matter are free to propagate
in the entire extra dimensional spacetime (the bulk). One of the
most popular braneworld models is the Rundall-Sundrom II setup [5].
In this model conventional 4D gravity can be recovered at large
scales (low energies) on a Minkowski braneworld embedded in a $5D$
Anti de Sitter (AdS) bulk. On the other hand, Dvali-Gabadadze and
Porrati (DGP) introduced a braneworld model which gravity is
modified at large distances rather than short distances in contrast
to other popular braneworld scenarios, because of an induced
four-dimensional Ricci scalar in the action on the brane [6]. This
term can be obtained by the quantum interaction between the matter
confined on the brane and the bulk gravitons. The DGP braneworld
scenario explains accelerated expansion of the universe via leakage
of gravity to extra dimension without need to introduce a dark
energy [8]. While the RS model produces ultra-violet (UV)
modification to the General Relativity, the DGP model leads to
infra-red (IR) modification of GR. However, by considering the
effect of an induced gravity term as a quantum correction in RS
model, we have a combined model that called "warped DGP braneworld"
in the literature [9]. This setup gives also a "self-accelerating"
phase in the brane evolution.

On the other hand, astrophysical data also indicate that $\omega$
lies in a very narrow strip close to $-1$. The case $\omega=-1$
corresponds to the cosmological constant. For $\omega$ less than
$-1$ the phantom dark energy is observed, and for $\omega$ more than
$-1$ (but less than$\frac{-1}{3}$) the dark energy is described by
quintessence. More ever, the analysis of the properties of dark
energy from recent observational data mildly favor models of dark
energy with $\omega$ crossing $-1$ line in the near past. So, the
phantom phase equation of state with $\omega < -1$ is still mildly
allowed by observations. In this case, the universe lives in its
phantom phase which ends eventually at a future singularity (Big
Rip). There are also a lot of evidence all around of a dynamical
equation of state, which has crossed the so called phantom divide
line $\omega=-1$ recently, at the value of red shift parameter
$z\approx 0.25$ [10]. Most of dark energy models treat scalar
field(s) as dark component(s) with a dynamical equation of state.
Currently scalar fields play crucial roles in modern cosmology. In
the inflationary scenario they generate an exponential rate of
evolution of the universe as well as density fluctuations due to
vacuum energy. It seems that the presence of a non-minimal coupling
(NMC) between scalar field and gravity is also necessary. There are
many theoretical evidences that suggest incorporation of an explicit
non-minimal coupling of scalar field and gravity in the action [11].
The nonzero non-minimal coupling arises from quantum corrections and
it is required also by the renormalization of the corresponding
field theory. Amazingly, it has been proven that the phantom divide
line crossing of dark energy described by a single minimally coupled
scalar field with general Lagrangian is even unstable with respect
to the cosmological perturbations realized on the trajectories of
the zero measure [12]. This fact has motivated a lot of attempts to
realize crossing of the phantom divide line by equation of state
parameter of scalar field as dark energy candidate in more
complicated frameworks. One of these attempts is a hybrid model,
composed of two scalar fields, quintessence and phantom, that
usually dubbed as quintom model in literature [13]. A quintom model
was initially proposed to obtain a model of dark energy with an EoS
parameter $\omega$ which satisfies $\omega > -1$ in the past and $
\omega < -1 $ at present. As we have emphasized, this model is
mildly favored by the current observational data fitting. Thus, the
quintom model is a dynamical scenario of dark energy with the
property that its EoS can smoothly cross over the cosmological
constant barrier $\omega=-1$. Recently, Zhang and Zhu have
considered a minimally coupled scalar field on the DGP braneworld
and realized crossing of the phantom divide line by considering two
possible cases [14]: for ordinary scalar field (quintessence) EoS of
dark energy crosses from $\omega>-1$ to $\omega<-1$ in negative
branch of DGP setup ( that is branch with $\epsilon=-1$) and for
phantom field EoS of dark energy crosses from $\omega<-1$ to
$\omega>-1$ in positive branch ($\epsilon=+1$) of DGP scenario. With
these preliminaries, in this paper we investigate crossing of the
phantom divide line by considering quintom field non-minimally
coupled to induced gravity on a warped DGP brane. Assuming a FRW
brane with induced gravity in the presence of a quintom field which
is non-minimally coupled to the induced Ricci scalar, we show that
the crossing occurs in a suitable range of model parameters. It is
surprising that in this case we have crossing of phantom divide in
both positive and negative branches ( $\varepsilon=\pm1$) of the
warped DGP brane. This phenomenon is depended on the choice of the
sign of non-minimal coupling parameter.

\section{A Dark Energy Model on the Warped DGP Brane}
\subsection{Warped DGP Braneworld}
We start with the action of the warped DGP model as follows
\begin{equation}
{\cal{S}}={\cal{S}}_{bulk}+{\cal{S}}_{brane},
\end{equation}
\begin{equation}
{\cal{S}}=\int_{bulk}d^{5}X\sqrt{-{}^{(5)}g}\bigg[\frac{1}{2\kappa_{5}^{2}}
{}^{(5)}R+{}^{(5)}{\cal{L}}_{m}\bigg]+\int_{brane}d^{4}x\sqrt{-g}\bigg[\frac{1}{\kappa_{5}^{2}}
K^{\pm}+{\cal{L}}_{brane}(g_{\alpha\beta},\psi)\bigg].
\end{equation}
Here ${\cal{S}}_{bulk}$ is the action of the bulk,
${\cal{S}}_{brane}$ is the action of the brane and ${\cal{S}}$ is
the total action. $X^{A}$ with $A=0,1,2,3,5$ are coordinates in bulk
while $x^{\mu}$ with $\mu=0,1,2,3$ are induced coordinates on the
brane. $\kappa_{5}^{2}$ is 5-dimensional gravitational constant.
${}^{(5)}R$ and ${}^{(5)}{\cal{L}}_{m}$ are 5-dimensional Ricci
scalar and matter Lagrangian respectively. $K^{\pm}$ is trace of
extrinsic curvature on either side of the brane.
${\cal{L}}_{brane}(g_{\alpha\beta},\psi)$  is the effective
4-dimensional Lagrangian. The action ${\cal{S}}$ is actually a
combination of Randall-Sundrum II and DGP model. In other words, an
induced curvature term is appeared on the brane in Randall-Sundrum
II model, hence the name {\it warped} DGP Braneworld [9]. Now
consider the brane Lagrangian as follows
\begin{equation}
{\cal{L}}_{brane}(g_{\alpha\beta},\psi)=\frac{\mu^2}{2}R-\lambda+L_{m},
\end{equation}
where $\mu$ is a mass parameter, $R$ is Ricci scalar of the brane,
$\lambda$ is tension of the brane and $L_{m}$ is Lagrangian of the
other matters localized on the brane. Assume that bulk contains only
a cosmological constant, $^{(5)}\Lambda$. With these choices, action
(1) gives either a generalized DGP or a generalized RS II model: it
gives DGP model if $\lambda=0$ and $^{(5)}\Lambda=0$, and gives RS
II model if $\mu=0$ [9]. The generalized Friedmann equation on the
brane is as follows
\begin{equation}
H^{2}+\frac{k}{a^{2}}=\frac{1}{3\mu^2}\bigg[\rho+\rho_{0}\Big(1+\varepsilon
{\cal{A}}(\rho,a)\Big)\bigg],
\end{equation}
where $\varepsilon=\pm 1$ is corresponding to two possible branches
of solutions ( two different embedding of the brane) in this warped
DGP model and
${\cal{A}}=\bigg[{\cal{A}}_{0}^{2}+\frac{2\eta}{\rho_{0}}
\Big(\rho-\mu^{2}\frac{{\cal{E}}_{0}}{a^{4}}\Big)\bigg]^{1/2}$ where
\,\, ${\cal{A}}_{0}\equiv
\bigg[1-2\eta\frac{\mu^{2}\Lambda}{\rho_{0}}\bigg]^{1/2}$,\,\, $\eta
\equiv\frac{6m_{5}^{6}}{\rho_{0}\mu^{2}}$\,\, with $0<\eta\leq1$
\,\,and \,\,$\rho_{0}\equiv
m_{\lambda}^{4}+6\frac{m_{5}^{6}}{\mu^{2}}$.  ${\cal{E}}_{0}$ is an
integration constant and corresponding term in the generalized
Friedmann equation is called dark radiation term. We neglect dark
radiation term in which follows. In this case, generalized Friedmann
equation (4) takes the following form
\begin{equation}
H^{2}+\frac{k}{a^2}=\frac{1}{3\mu^2}\bigg[\rho+\rho_{0}+\varepsilon
\rho_{0}\Big({\cal{A}}_{0}^{2}+\frac{2\eta\rho}{\rho_{0}}\Big)^{1/2}\bigg],
\end{equation}
where $\rho$ is the total energy density, including scalar fields
and dust matter energy densities on the brane:
\begin{equation}
\rho=\rho_\varphi+\rho_\sigma+\rho_{dm}
\end{equation}
and $\rho_{0}$ is given by,
\begin{equation}
\rho_{0}=\frac{6\mu^2}{r_{c}^2},
\end{equation}
where the crossover radius is defined as $r_{c}= \kappa_{5}^{2}
\mu^{2}$.

\subsection{A Quintom Dark Energy Model on the Warped DGP Brane}
Now we consider two scalar fields, one quintessence and the other
phantom field both non-minimally coupled to induced gravity on the
warped DGP brane. The action of this non-minimal quintom model is
given by
\begin{equation}
{\cal{S}}_{quint}=\int_{brane}d^{4}x\sqrt{-g}\Big[\frac{1}{2}\xi
R(\varphi^{2}+\sigma^{2})-\frac{1}{2}\partial_{\mu}\varphi\partial^{\mu}\varphi+
\frac{1}{2}\partial_{\mu}\sigma\partial^{\mu}\sigma-V_{1}(\varphi)-V_{2}(\sigma)\Big],
\end{equation}
where $\xi $ is a non-minimal coupling, $R$ is Ricci scalar of the
brane, $\varphi$ and $ \sigma $ are the normal (quintessence) and
phantom field respectively and $V_{1}(\varphi)$ and $V_{2}(\sigma)$
are corresponding potentials. We have assumed a conformal coupling
of the scalar fields and induced gravity and these fields play the
role of quintom dark-energy component on the brane. Variation of the
action with respect to $\varphi$ gives the equation of motion of the
normal scalar field
\begin{equation}
\ddot{\varphi}+3H\dot{\varphi}-\xi R\varphi
+\frac{dV_{1}}{d\varphi}=0.
\end{equation}
and variation of the action with respect to $\sigma$ gives the
equation of motion of phantom field
\begin{equation}
\ddot{\sigma}+3H\dot{\sigma}+\xi R\sigma -\frac{dV_{2}}{d\sigma}=0.
\end{equation}
The energy density and pressure of quintom are given by the
following relation respectively
\begin{equation}
\rho_{quint}=\rho_\varphi+\rho_\sigma=\frac{1}{2}(\dot{\varphi}^{2}-\dot{\sigma}^{2})+V_{1}(\varphi)+V_{2}(\sigma)-6\xi
H(\varphi\dot{\varphi}+\sigma\dot{\sigma})-3\xi
H^{2}(\varphi^{2}+\sigma^{2})
\end{equation}
and
$$p_{quint}=p_\varphi+p_\sigma=\frac{1}{2}(\dot{\varphi}^{2}-
\dot{\sigma}^{2})-V_{1}(\varphi)-V_{2}(\sigma)+2\xi\Big(\varphi\ddot{\varphi}+2\varphi
H\dot{\varphi}+\dot{\varphi}^{2}+\sigma\ddot{\sigma}+2\sigma
H\dot{\sigma}+\dot{\sigma}^{2}\Big)$$
\begin{equation}
+\xi(2\dot{H}+3H^2)(\varphi^2+\sigma^2)
\end{equation}
In which follows, by comparing the modified Friedmann equation in
the warped DGP braneworld with the standard Friedmann equation, we
deduce the equation of state of quintom field. This is reasonable
since all observed features of dark energy are essentially derivable
in general relativity [14,15]. The standard Friedmann equation in
four dimensions is written as
\begin{equation}
H^2+\frac{k}{a^2}=\frac{1}{3\mu^2}(\rho_{dm}+\rho_{de}),
\end{equation}
where $\rho_{dm}$ is the dust matter density, while $\rho_{de}$ is
dark energy density. Comparing this equation with equation (5) we
find
\begin{equation}
\rho_{de}=\rho_{\varphi}+\rho_{\sigma}+\rho_{0}+\varepsilon\rho_0\Big(A_0^2+2\eta\frac{\rho}{\rho_0}\Big)^{\frac{1}{2}}.
\end{equation}
Non-minimal coupling of the scalar fields and Ricci curvature on the
brane doesn't break the validity of conservation of the scalar
fields effective energy density
\begin{equation}
\frac{d\rho_{quint}}{dt}+3H(\rho_{quint}+p_{quint})=0.
\end{equation}
Since the dust matter obeys the continuity equation and the Bianchi
identity keeps valid, dark energy itself satisfies the continuity
equation
\begin{equation}
\frac{d\rho_{de}}{dt}+3H(\rho_{de}+p_{de})=0
\end{equation}
where $p_{de}$ denotes the pressure of the dark energy. The equation
of state for the dark energy can be written as follows
\begin{equation}
w_{de}=\frac{p_{de}}{\rho_{de}}=-1+\frac{1}{3}\frac{d\ln\rho_{de}}{d\ln(1+z)}.
\end{equation}
Using equations (14) and (16) we find
$$\frac{d\ln\rho_{de}}{d\ln(1+z)}=\frac{3}{\rho_{de}}\bigg[\rho_{\varphi}+p_{\varphi}+\rho_{\sigma}+p_{\sigma}
$$
\begin{equation}
+\varepsilon\eta\bigg(A_0^2+2\eta\frac{\rho_{\varphi}+\rho_{\sigma}+\rho_{dm}}{\rho_0}\bigg)^{-\frac{1}{2}}\bigg(\rho_{\varphi}
+p_{\varphi}+\rho_{\sigma}+p_{\sigma}+\rho_{dm}\bigg)\bigg].
\end{equation}
There are three possible cases in this setup: if
$\Big(\frac{1}{3}\frac{d\ln\rho_{de}}{d\ln(1+z)}\Big)>0$, we have a
quintessence model; if
$\Big(\frac{1}{3}\frac{d\ln\rho_{de}}{d\ln(1+z)}\Big)<0$ the model
is phantom and if
$\Big(\frac{1}{3}\frac{d\ln\rho_{de}}{d\ln(1+z)}\Big)=0$, the dark
component is a cosmological constant. Evidently, in this setup
non-minimal coupling of scalar fields and induced gravity plays a
crucial role supporting or preventing phantom divide line crossing.
In this respect, the differences between the minimal and non-minimal
setups will be more clear if we write the explicit dynamics of
equation of state parameter. On the other hand, the effect of warp
factor which appears in the definition of $\rho_{de}$, will be
highlighted in forthcoming arguments. We choose the following
exponential potential with motivation that this type of potential
can be solved exactly in the standard model
\begin{equation}
V_{1}(\varphi)=V_{01} \exp(-\lambda_{1}\frac{\varphi}{\mu}),
\end{equation}
and
\begin{equation}
V_{2}(\sigma)=V_{02} \exp(-\lambda_{2}\frac{\sigma}{\mu}),
\end{equation}
where $V_{01}$, $V_{02}$, $\lambda_{1}$, $\lambda_{2}$ and $\mu$ are
constant. Therefore, we have
$$\omega=-1+\frac{1}{\rho_{de}}\Bigg[(\dot{\varphi}^2-\dot{\sigma}^2)+2\xi\Big(-H(\varphi\dot{\varphi}+\sigma\dot{\sigma})
+\dot{H}(\varphi^2+\sigma^2)+\varphi\ddot{\varphi}+\sigma\ddot{\sigma}+\dot{\varphi}^2+\dot{\sigma}^2\Big)
$$
$$+\Big[(\dot{\varphi}^2-\dot{\sigma}^2)+2\xi\Big(-H(\varphi\dot{\varphi}+\sigma\dot{\sigma})
+\dot{H}(\varphi^2+\sigma^2)+\varphi\ddot{\varphi}+\sigma\ddot{\sigma}+\dot{\varphi}^2+\dot{\sigma}^2\Big)+\rho_{dm}\Big]$$
\begin{equation}
\Big[\varepsilon\eta\Big(A_{0}^2+2\eta\frac{\frac{1}{2}(\dot{\varphi}^{2}-\dot{\sigma}^{2})+V_{1}(\varphi)+V_{2}(\sigma)-6\xi
H(\varphi\dot{\varphi}+\sigma\dot{\sigma})-3\xi
H^{2}(\varphi^{2}+\sigma^{2})+\rho_{dm}}{\rho_0}\Big)^{-\frac{1}{2}}\Big]\Bigg]
\end{equation}
As a comparison, in the minimal case (with $\xi=0$) and neglecting
the warp effect, when we consider just a quintessence field and
choosing the sign of $\varepsilon$ to be negative, two remaining
terms on the right hand side of equation (21) will have opposite
signs and the EoS parameter essentially crosses the phantom divide
line [14]. However, in our non-minimal quintom model the situation
is more complicated and it is not simple to conclude that there is
crossing of the phantom divide line or not just by defining
$\varepsilon$ sign since non-minimal coupling itself plays a crucial
role in this case. We consider $\xi$ (the non-minimal coupling of
scalar fields and induced gravity) as a fine-tuning parameter in
this setup. In which follows, we use some parameters like $\Omega_m$
( the value of energy density of dust matter over the critical
density defined as $\rho_{c}= 24\mu^{2}H_{0}^{2}$\,\,),
$\Omega_{ki}$ ( present value of kinetic energy density of the
scalar field over the critical density ), $\Omega_{rc}$ ( present
value of the energy density of $\rho_0$ over the critical density)
and non-minimal coupling to calculate EoS parameter, $\omega(z)$. If
we change the values of these parameters in appropriate manner (
subjected to observational constraints), the redshift at which
crossing of the phantom divide line occurs will change since it is a
model dependent quantity in this respect. We explain further these
behaviors of phantom divide line crossing in our forthcoming
arguments and within a numerical scheme. For these numerical
calculations we introduce a new parameter defined as $s=-\ln (1+z)$
and in all figures the behavior of $\omega$ is plotted versus $s$.
In this numerical calculations, we need explicit form of
$\ddot{\varphi}$ (or $\ddot{\sigma}$) in terms of other quantities.
For instance $\ddot{\varphi}$ is given as follows
$$\ddot{\varphi}=\frac{6\xi\varphi\big[(\dot{\varphi}^2-\dot{\sigma}^2)+
2\xi\big(-H(\varphi\dot{\varphi}+\sigma\dot{\sigma})+\dot{\varphi}^2+
\dot{\sigma}^2\big)+\rho_{dm}\big]\Big(\frac{b}{-2\big(\mu^{2}+\xi(\varphi^2+\sigma^2)b\big)}\Big)}
{1+(\varphi-\frac{\sigma^2}{\varphi})\frac{6\xi^2\varphi\dot{\varphi}
b}{\mu^2+\xi(\varphi^2+\sigma^2)b}}$$
\begin{equation}
+\frac{\frac{\lambda_{1}}{\mu}V_{1}(\varphi)+[\frac{\sigma
x}{\varphi\dot{\varphi}}+\frac{y}{\dot{\sigma}}]
\frac{3\xi\varphi\sigma
b}{(\mu^2+\xi(\varphi^2+\sigma^2)b)}-3H(\dot\varphi-4\xi\varphi
H)}{1+(\varphi-\frac{\sigma^2}{\varphi})\frac{6\xi^2\varphi\dot{\varphi}b}{\mu^2+\xi(\varphi^2+\sigma^2)b}}
\end{equation}
where we have defined
\begin{equation}
x\equiv\frac{3H(\dot{\varphi}^{2}-4\xi \varphi
H\dot{\varphi})+\dot{V_{1}}(\varphi)}{6\xi\varphi\dot{\varphi}}
\end{equation}
and
\begin{equation}
y\equiv\frac{3H(\dot{\sigma}^{2}+4\xi \sigma
H\dot{\sigma})+\dot{V_{2}}(\sigma)}{6\xi\sigma\dot{\sigma}}
\end{equation}
and
\begin{equation}
b\equiv1+\varepsilon\eta(A_0^2+2\eta\frac{\rho}{\rho_0})^{-\frac{1}{2}}
\end{equation}
On the other hand, Friedmann equation given by (5) now takes the
following complicated form
$$ (\mu^2+g)^{2}H^4+2f(3\mu^2+g)H^3+\Big[f^2-2l(3\mu^2+g)+2\eta\varepsilon^2\rho_0 g\Big]H^2$$
\begin{equation}
+\Big(-2fl+\varepsilon^2\rho_0\eta
f\Big)H-2\eta\varepsilon^2\rho_0(l-\rho_0)- \varepsilon^2\rho_0^2
A_{0}^2+l^2=0
\end{equation}

 where
\begin{equation}
g\equiv3\xi H^{2}(\varphi^{2}+\sigma^{2}),
\end{equation}
\begin{equation}
l\equiv\frac{1}{2}
(\dot{\varphi^2}-\dot{\sigma^2})+V_{1}(\varphi)+V_{2}(\sigma)+\rho_{dm}+\rho_{0},
\end{equation}
and
\begin{equation}
f\equiv6\xi H(\varphi\dot{\varphi}+\sigma\dot{\sigma}).
\end{equation}
Equation (26) is a quadratic equation in terms of $H$ and in
principle has four roots. We show these roots as $h_{1}$, $h_{2}$,
$h_{3}$ and $h_{4}$. Two of these roots, say, $h_{1}$ and $h_{2}$
are negative, excluded from observational ground. The other two
roots, $h_{3}$ and $h_{4}$, are positive and as we will show have
the capability to account for phantom divide line crossing. Since
behaviors of these roots in treating phantom divide line crossing
are very similar, in which follows we consider only one of these
roots, say $h_{4}$,  and investigate its cosmological consequences.
The effect of warp factor on the dynamics of these solutions will be
discussed later. In table $1$, we have obtained some reliable ranges
of non-minimal coupling to have crossing of the phantom divide line
in this setup. We have assumed the age of universe to be $13\,Gyr$
and with this choice the values of $\xi$ are constraint to the
ranges shown in the table. On the other hand, observational data
show that crossing of the phantom divide line is occurred in
redshift $z\simeq 0.25$ ( though a model dependent value but
suitable for our purposes) so we have obtained the value of $\xi$
which is corresponding to this value in last column of the table.
The results of numerical calculations are shown in figure $1$ for
two branches of this DGP-inspired model and with different values of
non-minimal coupling $\xi$. In this analysis the best ranges of
values for $\xi$ to have a reliable model in comparison with
observational data are obtained. Note that in all of our numerical
calculations we have assumed $\Omega_{ki}=0.01$, $\Omega_{rc}=0.01$,
$\Omega_{m}=0.3$ and $\lambda_{1}$ and $\lambda_{2}=0.5$, and also
for investigating the effect of non-minimal coupling we have set
$A_0=1$, $H_0=1$, $\mu=1$ and $\eta=0.99$. Figure $1$ shows the
crossing of phantom divide line obtains in both positive and
negative branches of the DGP-inspired model. However, there is a
difference between crossing behavior of these solutions. Indeed, as
it could be understood from the figures, for positive branch of the
model there is crossing of the PDL (from phantom phase $\omega<-1$
to quintessence phase $\omega>-1$) only for positive non-minimal
coupling parameter. By contrast, the above result for negative
branch of DGP model holds only for negative values of NMC parameter
while EoS of dark energy has transition from quintessence phase
$\omega>-1$ to phantom phase $ \omega<-1$.

\begin{table}
\begin{center}
\caption{Acceptable range of $\xi$ ( constraint by the age of the
universe) for $h_{4}$ ( a positive root of $H$ as given by equation
(26)) to have crossing of the phantom divide line. } \vspace{0.5 cm}
\begin{tabular}{|c|c|c|c|c|c|c|c|}
  \hline
  \hline $\varepsilon$&$\xi$ &Acceptable range of $\xi$ & The value of $\xi$ for z=0.25  \\
  \hline +1&positive & $0.088\leq\xi\leq0.217$ &0.126  \\
  \hline +1&negative & no crossing & --- \\
 \hline -1&positive &no crossing &--- \\
  \hline-1&negative &$-0.292\leq\xi\leq-0.215$&-0.251  \\
 \hline
\end{tabular}
\end{center}
\end{table}

\begin{figure}[htp]
\begin{center}
\includegraphics{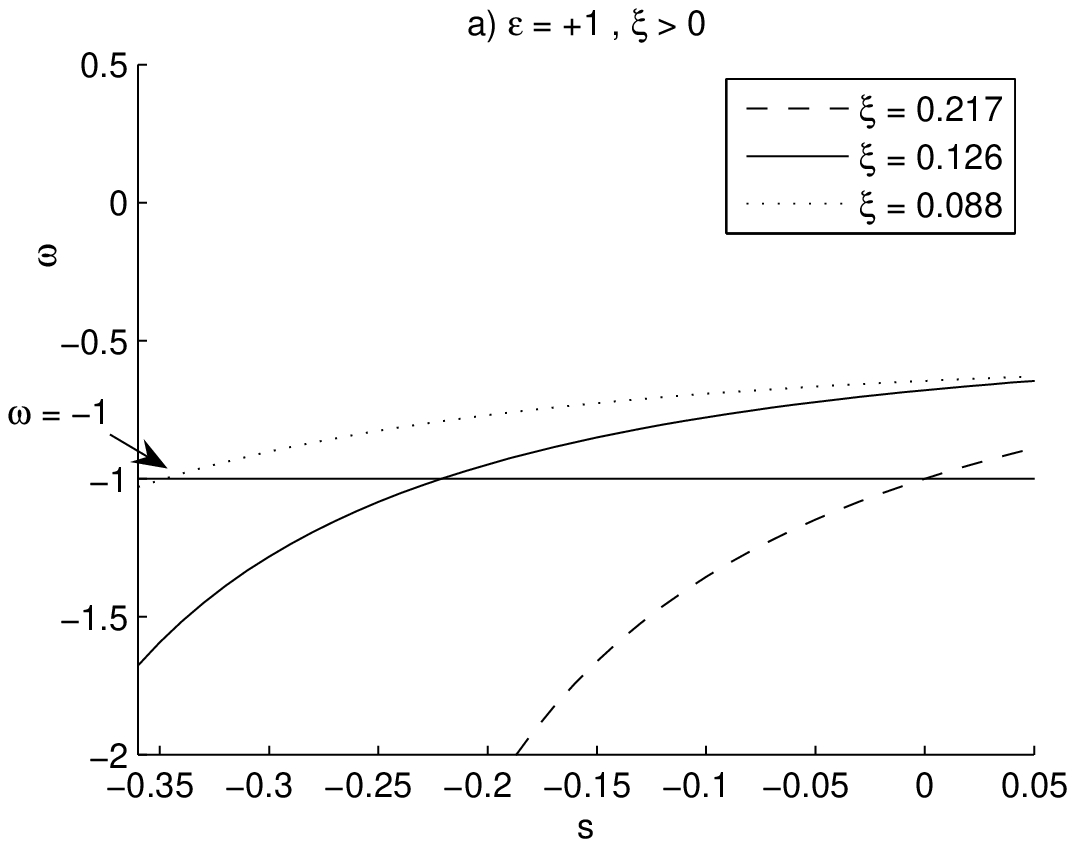} \vspace{5cm}\includegraphics{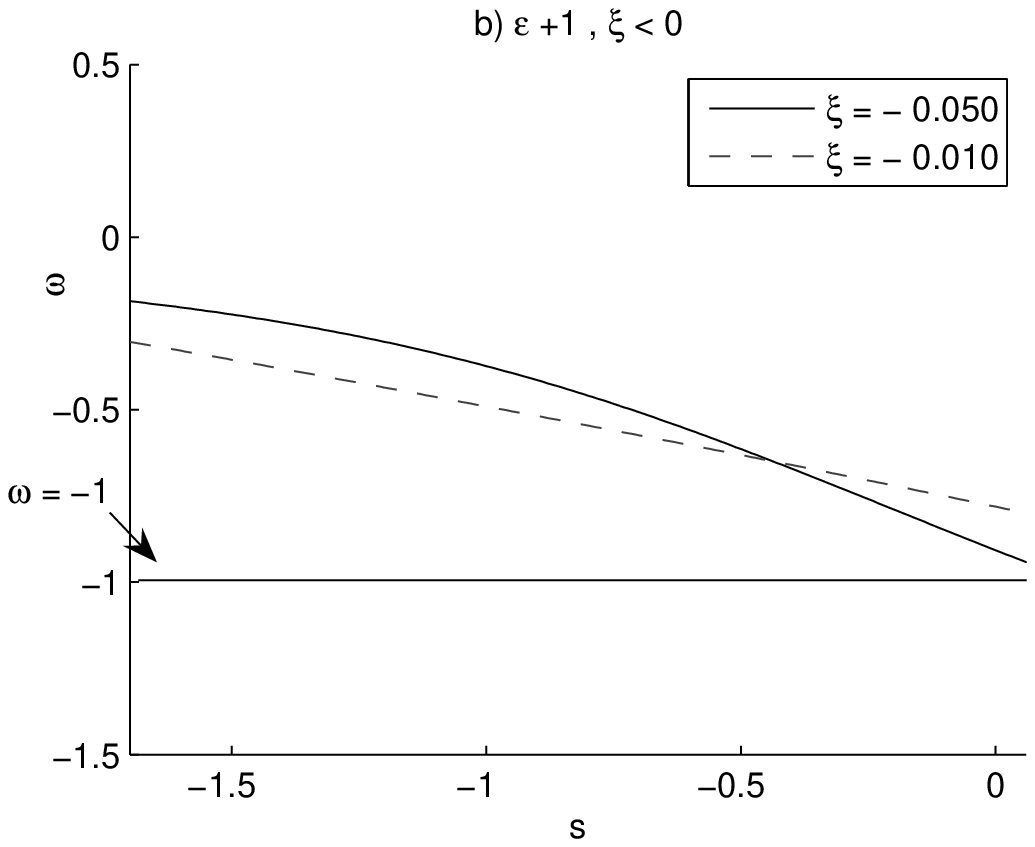}\vspace{5cm}\includegraphics{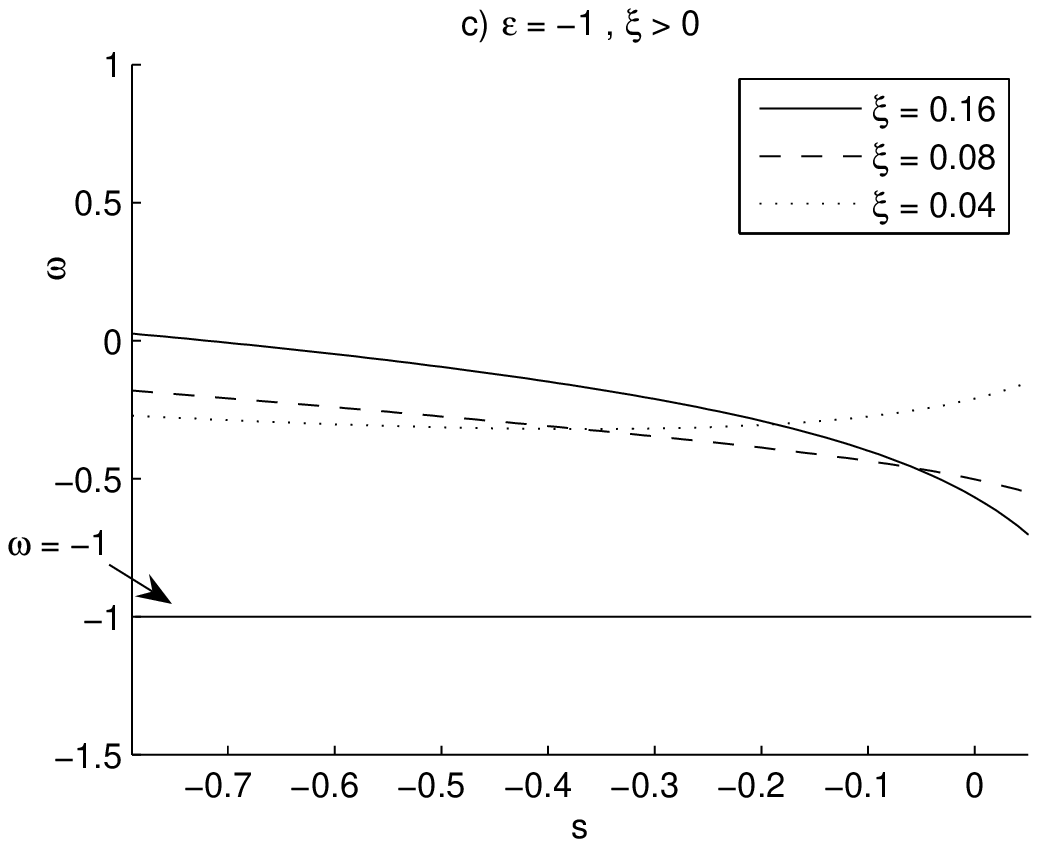}
\vspace{5cm}\includegraphics{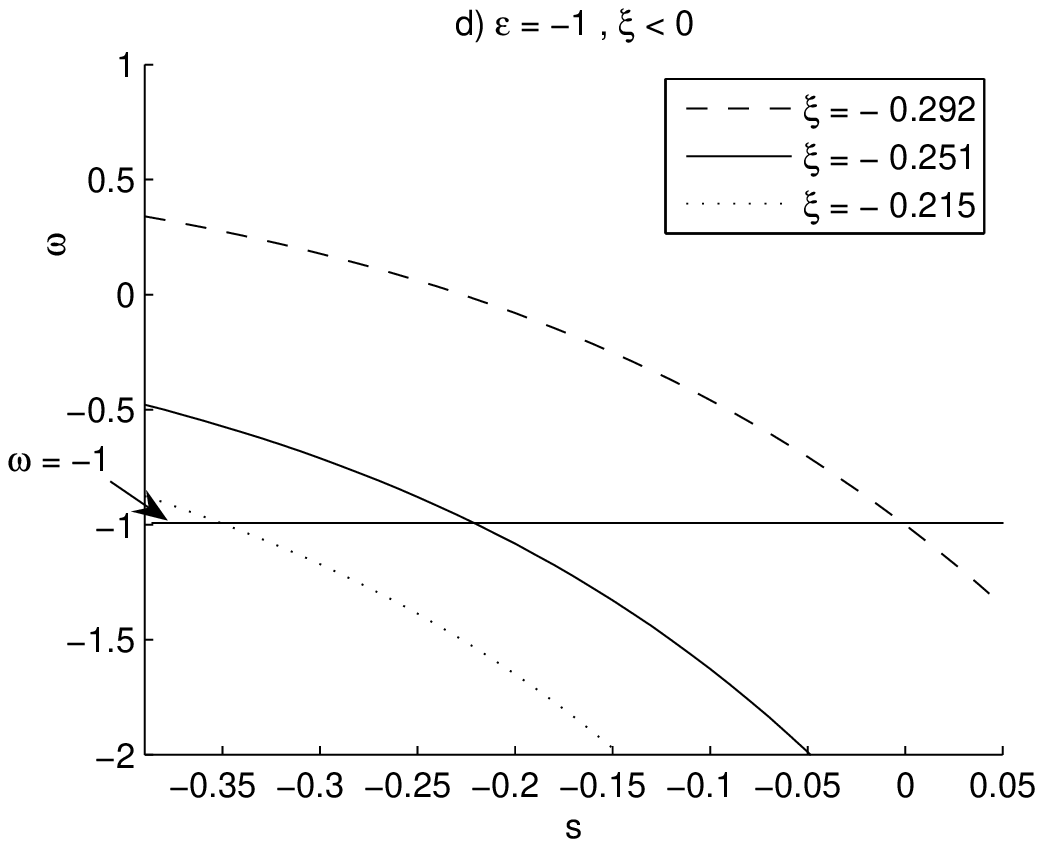}\vspace{1cm}
\end{center}
\vspace{-2 cm}
 \caption{\small {a) For positive branch ($\varepsilon=+1$) of this DGP-inspired scenario we find: a) with
 positive nonminimal coupling the EoS of dark energy crosses $\omega=-1$ ( from phantom phase to quintessence) for
 $\xi=0.126$ at $s=-0.22$ or $z=0.25$,  b) There is no crossing of the phantom divide
 line with negative values of non-minimal coupling. For negative
 branch (that is, $\varepsilon=-1$) of the scenario we see that:
 c) There is no crossing of the phantom divide line
 with positive values of non-minimal coupling, but, d) with negative nonminimal coupling
 the EoS of dark energy crosses $\omega=-1$ ( from quintessence to phantom phase) for
 $\xi=-0.251$ at $s=-0.22$ or $z=0.25$.
 This case has very good agreement with observational data that favor crossing from quintessence to phantom phase[16,17]. }}
\end{figure}

In figure $2$ we have plotted the behavior of deceleration parameter
$q$ versus $s$ in both branches of DGP-inspired scenario. We see
that in positive branch of the scenario and with positive values of
non-minimal coupling, deceleration parameter vanishes at a moment in
future with redshift $z\approx-0.37$ for $\xi=0.126$. On the other
hand, in negative branch by decreasing the values of $\xi$ (which is
negative in this case), the deceleration parameter vanishes at late
time epoches of the universe evolution. As an important result, by
incorporating the non-minimal coupling we have an accelerated
behavior even in the negative branch ( $\varepsilon=-1$) of this
DGP-inspired scenario.

\begin{figure}[htp]
\begin{center}
\includegraphics{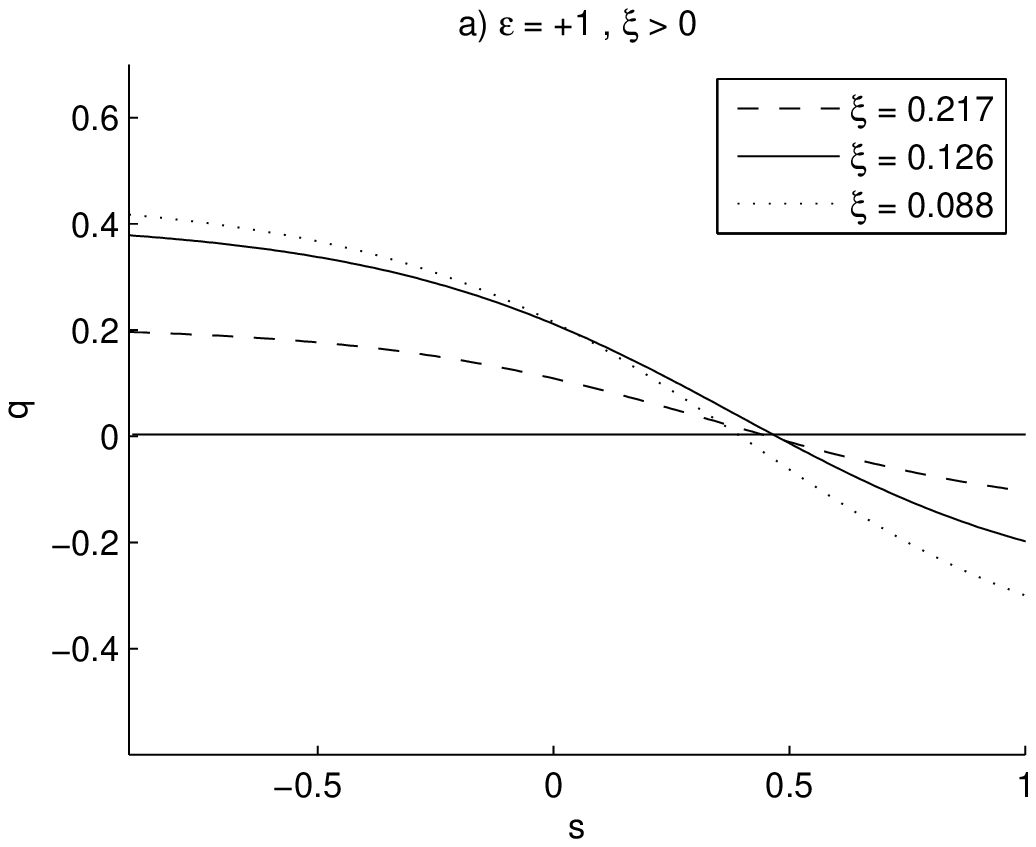} \vspace{5cm}\includegraphics{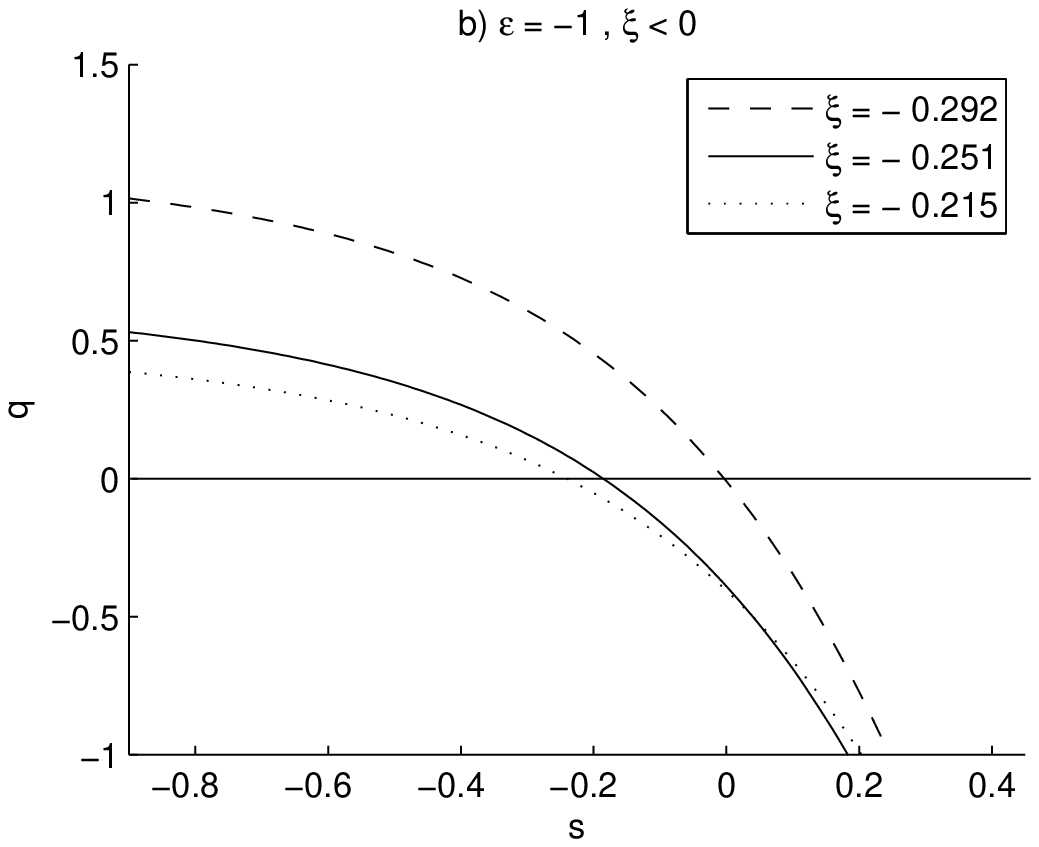}\vspace{5cm}
\end{center}
\vspace{2.5cm}
 \caption{\small {a) deceleration parameter in $DGP^{(+)}$ branch of the model which
 vanishes for instance at $s\approx0.45 $ for $\xi=0.217$. b) deceleration parameter in $DGP^{(-)}$ branch
 of the model which vanishes for instance at $s\approx0$ for $\xi=-0.292$. This vanishing shows the transition to
 acceleration or deceleration phase of the cosmological dynamics.}}
\end{figure}

In figure $3$ we show the behavior of the EoS parameter of dark
energy with different values of the parameter $\eta$ which is
related to warp effect. It is clearly seen that for sufficiently
small values of $\eta$, in both branches of the model EoS parameter
crosses the cosmological constant line in relatively small values of
redshifts. Both of two possible crossing; from phantom phase to
quintessence and from quintessence phase to phantom are possible in
this scenario.

\begin{figure}[htp]
\begin{center}
\includegraphics{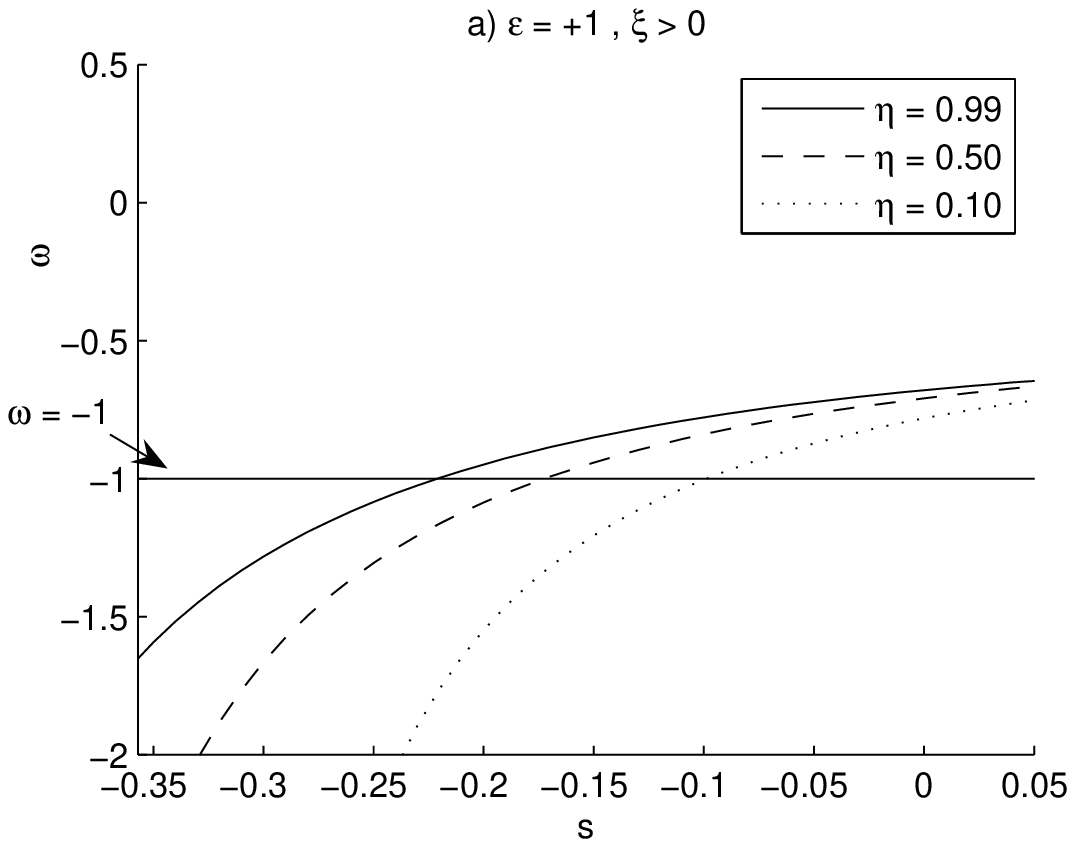} \vspace{5cm}\includegraphics{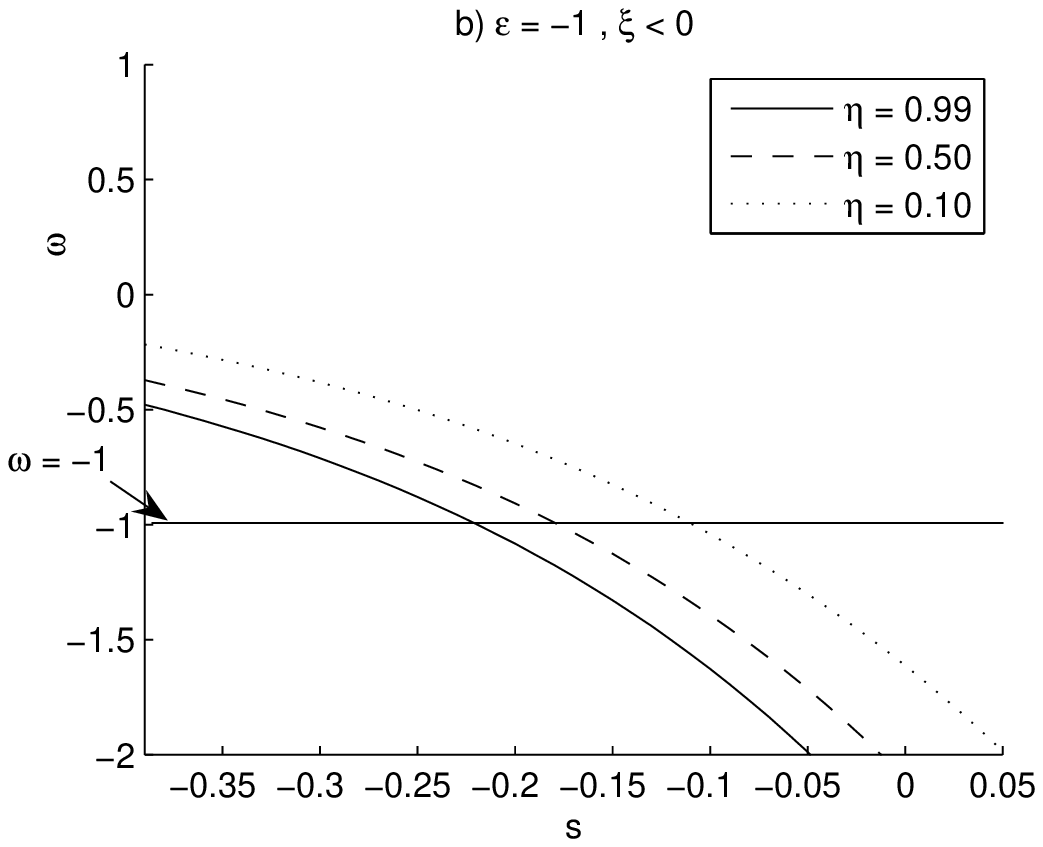}\vspace{5cm}
\end{center}
\vspace{2.5cm} \caption{\small {The role played by parameter $\eta$
which is related to warp effect on the crossing of the phantom
divide line. For sufficiently small values of $\eta$, equation of
state parameter, $\omega$, crosses the phantom divide line in
relatively small values of redshifts. a) In $DGP^{(+)}$ branch of
the model and with $\xi=0.126$, the EoS of dark energy crosses
$\omega=-1$ line for $\eta=0.99$ at about $s=-0.22$. For $\eta=0.5$
and $z=0.25 $ there is a crossing at $s\approx-0.17$ and so on. b)In
$DGP^{(-)}$ branch of the model and with $\xi= -0.251$, the EoS of
dark energy crosses $\omega=-1$ line for $\eta=0.99$ at about
$s=-0.22$, for instance.}}
\end{figure}
Figure $4$ is a three dimensional plot of the parameter $\omega$
versus $s$ and $\eta$. This figure confirm our previous discussions
especially the role played by warp factor.
\begin{figure}[htp]
\begin{center}
\includegraphics{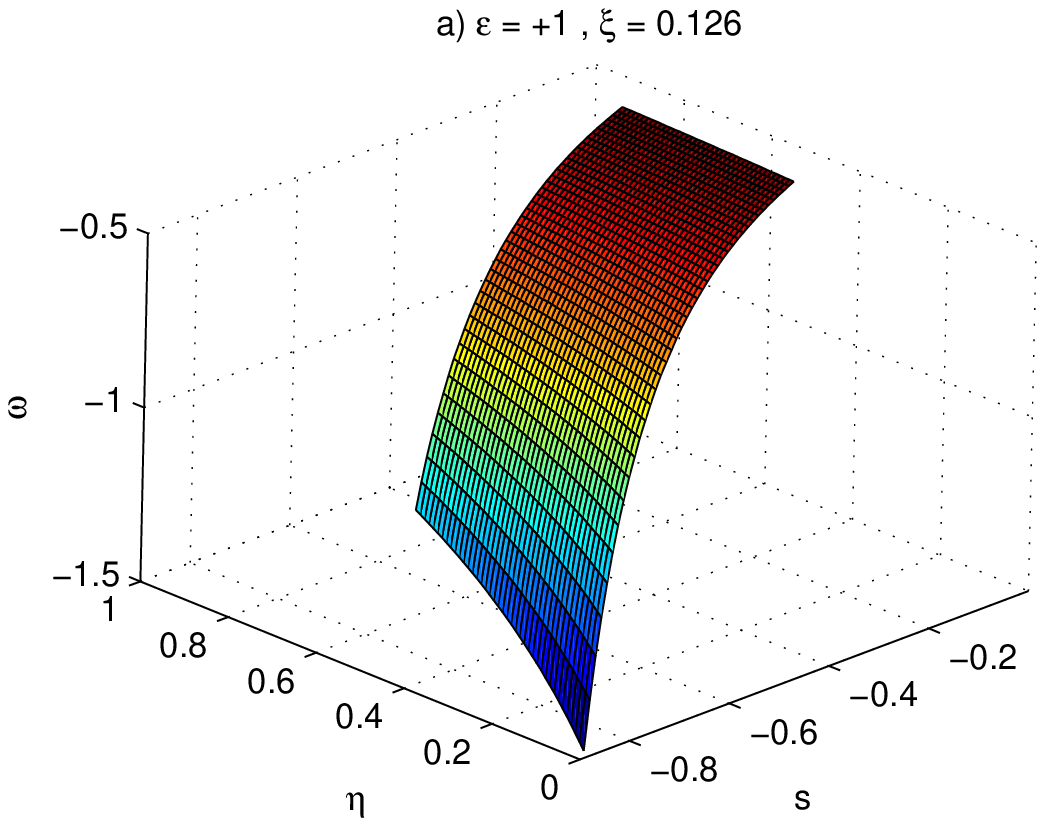} \vspace{5cm}\includegraphics{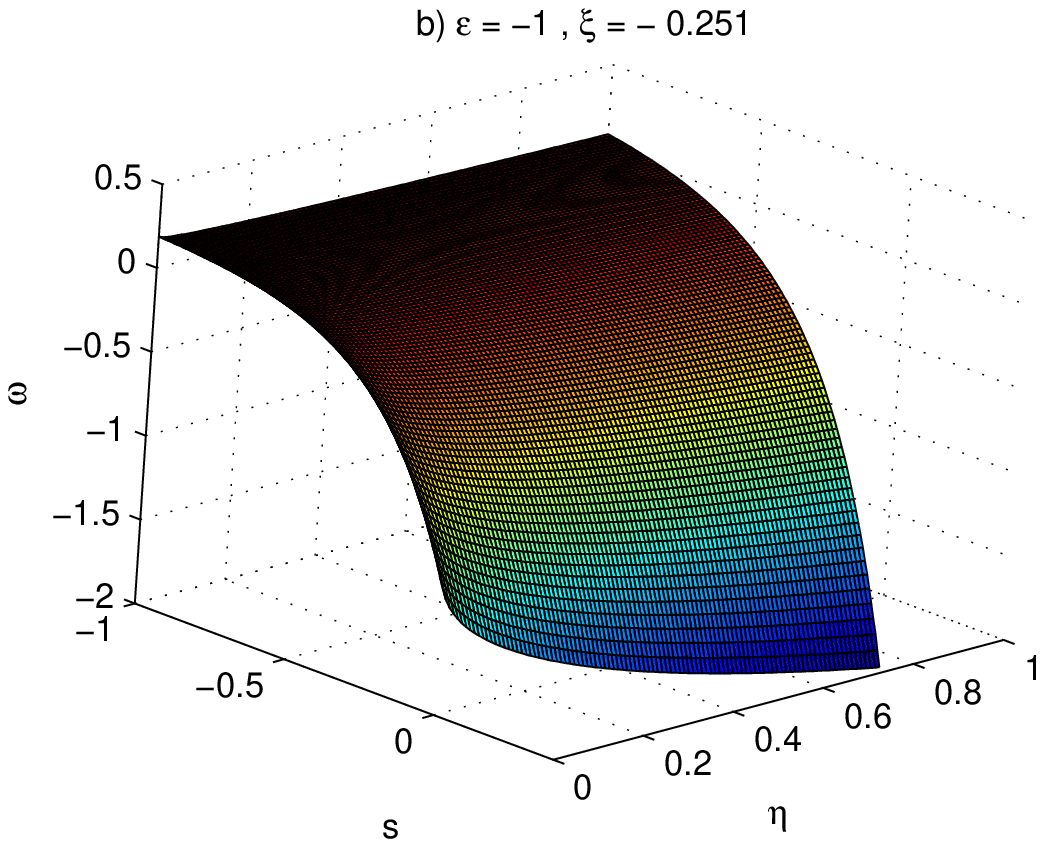}\vspace{5cm}
\end{center}
\vspace{0cm} \caption{\small {Three dimensional plot of the phantom
divide line crossing: a) In DGP$^{(+)}$ branch of the model with
$\xi=0.126$ and b) In DGP$^{(-)}$ branch of the model with
$\xi=-0.251$. Cosmological line crossing is possible in both of
these branches. }}
\end{figure}

\newpage
Now we discuss two especial cases of our model separately: a
quintessence and a phantom phase.

\subsection{Quintessence field}

In this subsection, we investigate dynamics of a quintessence field
non-minimally coupled to induced gravity on the background of a
warped DGP brane. The action of the model in this case takes the
following form
\begin{equation}
{\cal{S}}_{\varphi}=\int_{brane}d^{4}x\sqrt{-g}\Big[\frac{1}{2}\xi
R\varphi^{2}-\frac{1}{2}\partial_{\mu}\varphi\partial^{\mu}\varphi-V_1(\varphi)\Big].
\end{equation}
The energy density and pressure of this quintessence field are given
by
\begin{equation}
\rho_{\varphi}=\frac{1}{2}\dot{\varphi}^{2}+V_1(\varphi)-6\xi
H\varphi\dot{\varphi}-3\xi H^{2}\varphi^{2},
\end{equation}
and
\begin{equation}
p_{\varphi}=\frac{1}{2}\dot{\varphi}^{2}-V_1(\varphi)+2\xi(\varphi\ddot{\varphi}+2\varphi
H\dot{\varphi}+\dot{\varphi}^{2})+\xi\varphi^2(2\dot{H}+3H^2).
\end{equation}
The equation of state parameter of this quintessence field takes the
following form
$$\omega=-1+\frac{1}{\rho_{de}}\Bigg[\dot{\varphi}^2+2\xi\Big(-H\varphi\dot{\varphi}
+\dot{H}\varphi^2+\varphi\ddot{\varphi}+\dot{\varphi}^2\Big)
+\Big[\dot{\varphi}^2+2\xi\Big(-H\varphi\dot{\varphi}
+\dot{H}\varphi^2+\varphi\ddot{\varphi}+\dot{\varphi}^2\Big)+\rho_{dm}\Big]$$
\begin{equation}
\Big[\varepsilon\eta\Big(A_{0}^2+2\eta\frac{\frac{1}{2}\dot{\varphi}^{2}+V_1(\varphi)-6\xi
H\varphi\dot{\varphi}-3\xi
H^{2}\varphi^{2}+\rho_{dm}}{\rho_0}\Big)^{-\frac{1}{2}}\Big]\Bigg],
\end{equation}
where
\begin{equation}
\ddot{\varphi}=\frac{6\xi\varphi\big[\dot{\varphi}^2+2\xi\big(-H
\varphi\dot{\varphi}+\dot{\varphi}^2\big)+\rho_{dm}\big]\Big(\frac{b}{-2\big(\mu^{2}+
\xi\varphi^2b\big)}\Big)-3H(\dot\varphi-4\xi\varphi
H)+\frac{\lambda_{1}}{\mu}V_1(\varphi)} {1+\frac{6\xi^2\varphi^2
b}{\mu^2+\xi\varphi^2 b}}.
\end{equation}

\begin{table}
\begin{center}
\caption{Acceptable ranges of $\xi$ ( constraint by the age of the
universe) for $h_{4}$ ( a positive root of $H$ as given by equation
(26)) to have crossing of the phantom divide line by the equation of
state parameter of single quintessence field. } \vspace{0.5 cm}
\begin{tabular}{|c|c|c|c|c|c|c|c|}
 \hline
 \hline $\varepsilon$&$\xi$ &Acceptable range of $\xi$ & The value of $\xi$ for z=0.25  \\
 \hline +1&positive & $0.242\leq\xi\leq0.399$ &0.298  \\
 \hline +1&negative & no crossing & --- \\
 \hline -1&positive & no crossing &--- \\
 \hline-1&negative &$-0.147\leq\xi\leq-0.09$&-0.11  \\
 \hline
\end{tabular}
\end{center}
\end{table}
Table $2$ shows some acceptable ranges of non-minimal coupling of
quintessence field and induced gravity on the warped DGP setup.
These values are constraint by the age of the universe for one root
of equation (26) to have crossing of the phantom divide line by the
equation of state parameter. Figure $5$ shows the dynamics of EoS
parameter of this scalar field on the warped DGP braneworld. As we
see, in positive branch of the model ( with $\varepsilon=+1$), this
parameter crosses the cosmological constant line if non-minimal
coupling is positive. The crossing on negative branch ( with
$\varepsilon=-1$) occurs only with negative values of non-minimal
coupling. In both cases, crossing runs from phantom phase (
$\omega<-1$) to quintessence phase ( $\omega>-1$). There is no
crossing of PDL in $DGP^{+}$ branch with negative sign of $\xi$ and
in $DGP^{-}$ branch with positive sign of $\xi$.

\begin{figure}[htp]
\begin{center}
\includegraphics{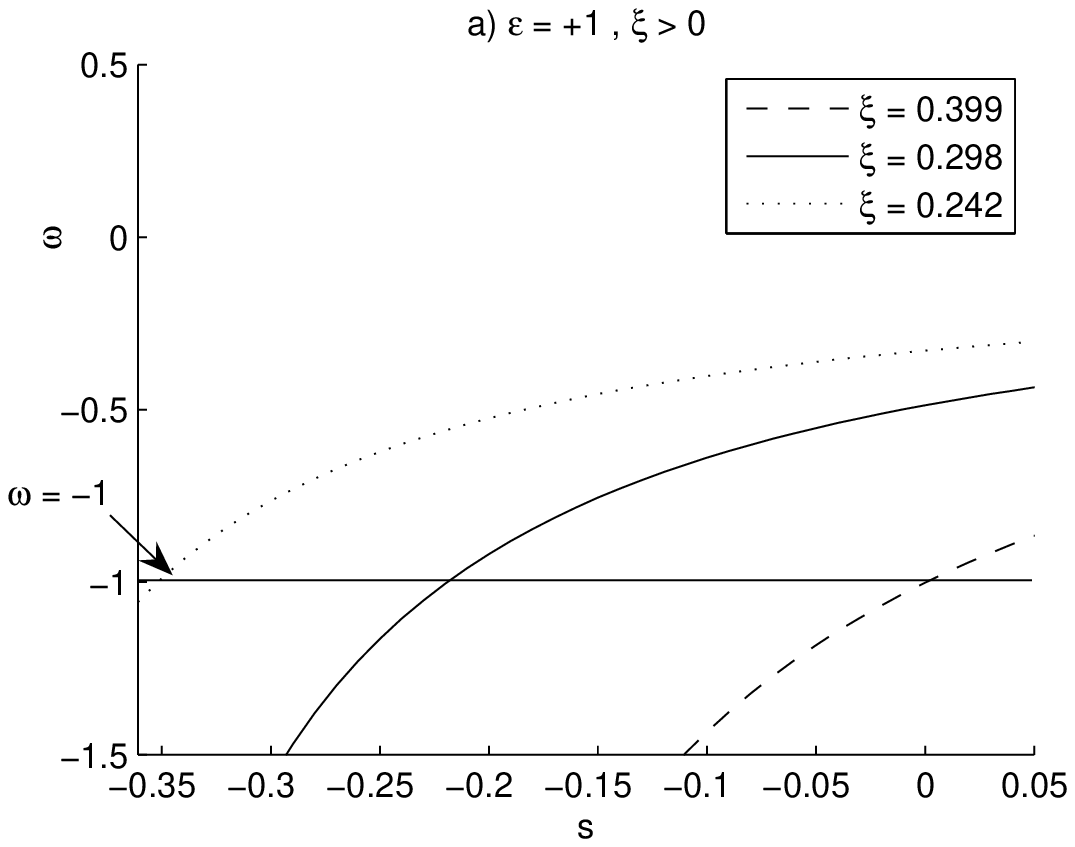} \vspace{5cm}\includegraphics{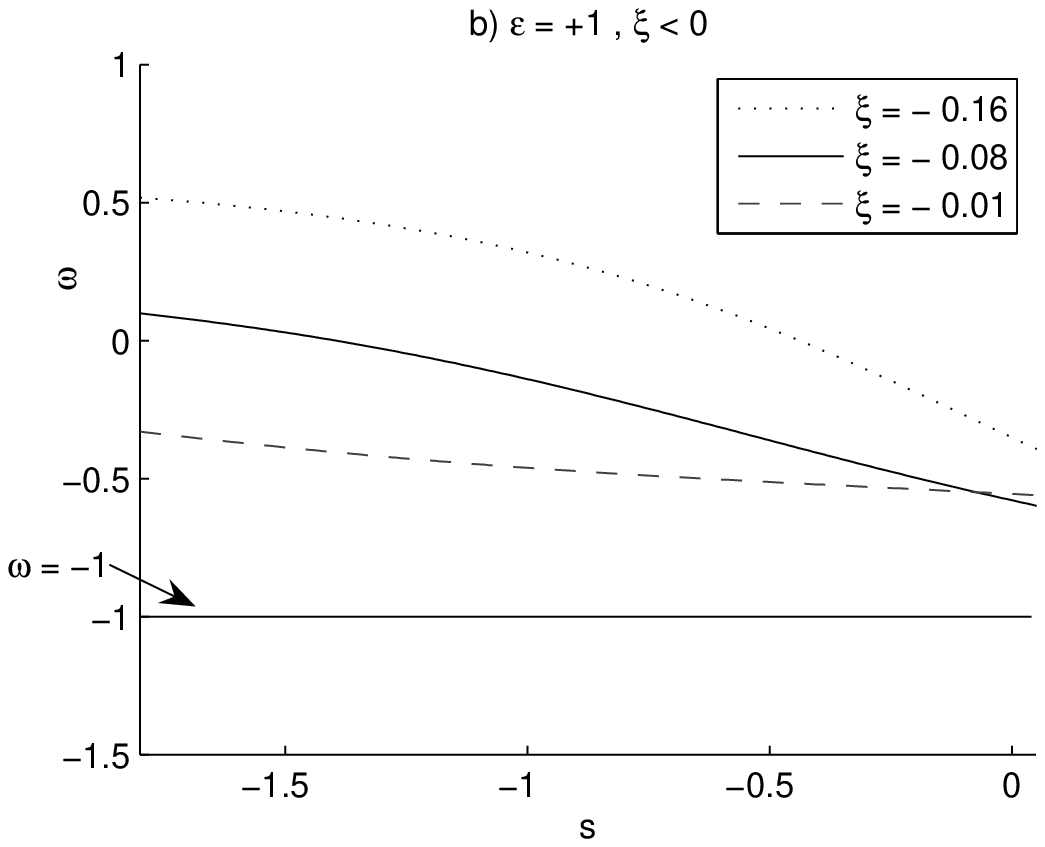}\vspace{5cm}\includegraphics{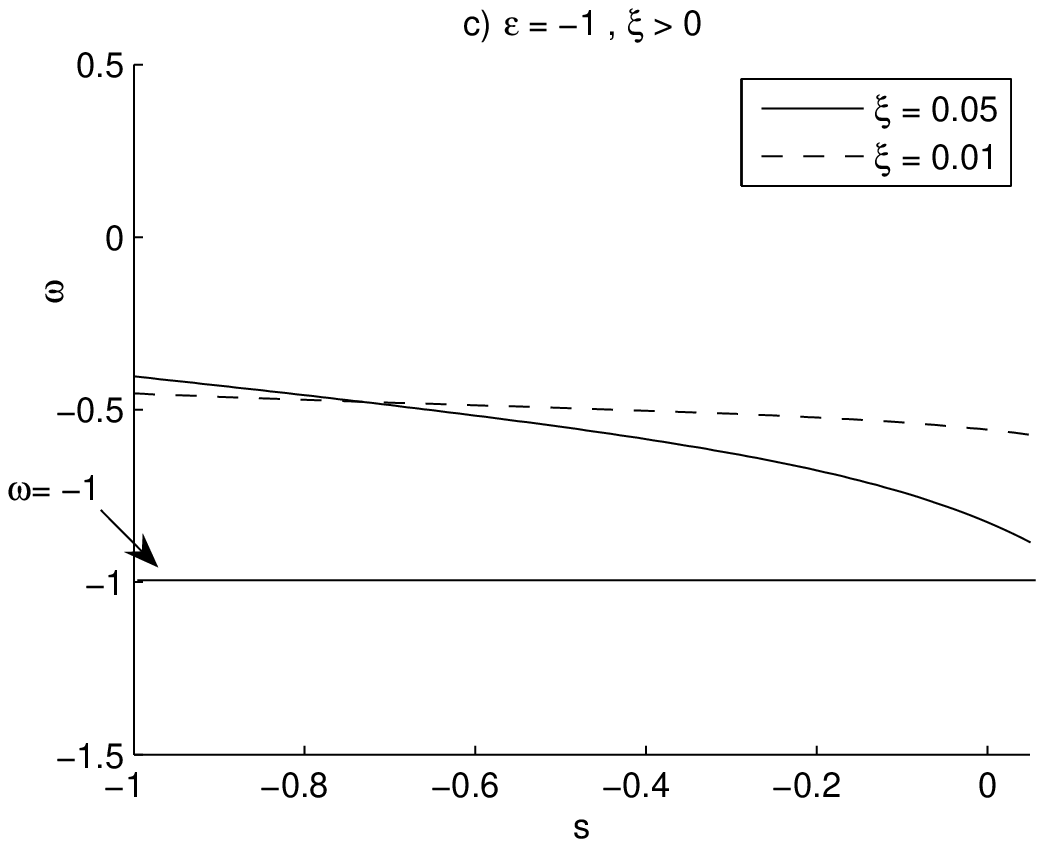}
\vspace{5cm}\includegraphics{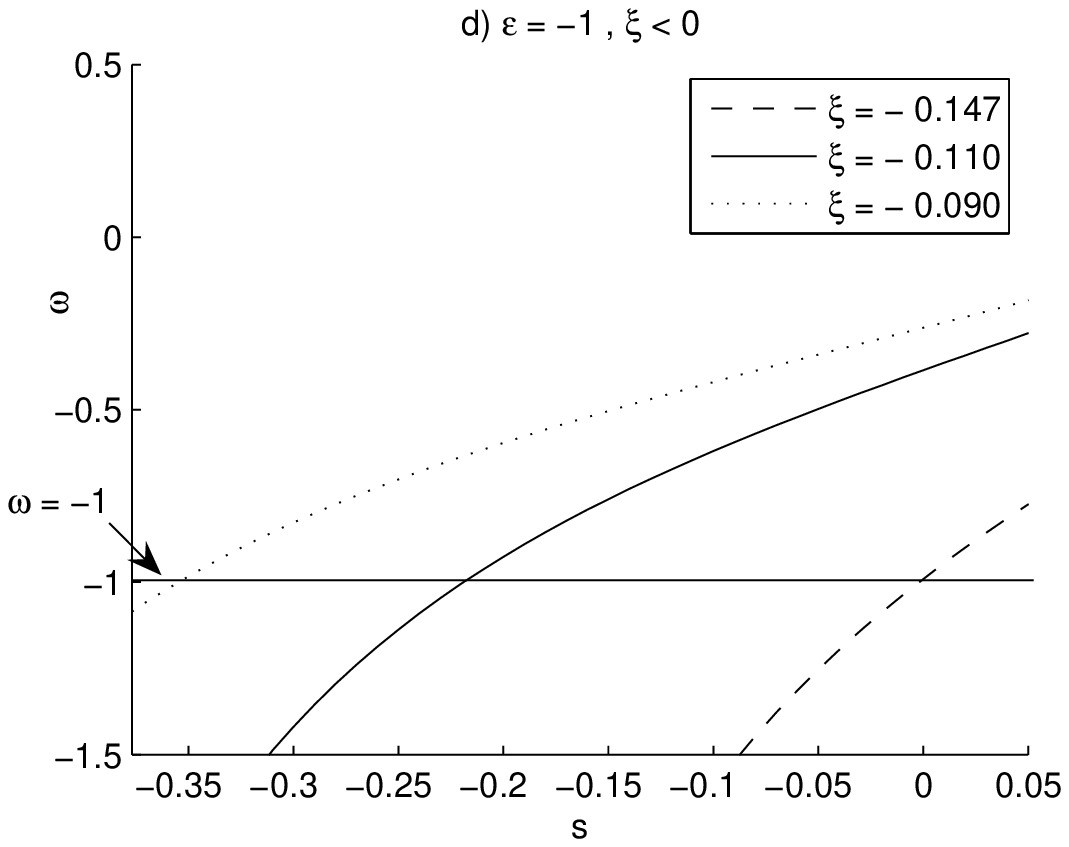}\vspace{1cm}
\end{center}
\vspace{-2 cm}
 \caption{\small { a) In positive branch of the scenario and with positive nonminimal
 coupling, $\omega$ crosses the cosmological constant line. For instance, if $\xi=0.298$
 this crossing occurs at $s=-0.22$ or $z=0.25$. b) There is no crossing of phantom divide
 line with negative values of non-minimal coupling $\xi$ in the positive branch of the model.
 c)In negative branch of the model with positive values of non-minimal
 coupling, $\omega$ never crosses the cosmological constant line. d) In negative branch and with
 negative nonminimal coupling the EoS of dark energy
 crosses $\omega=-1$ line. For instance, if $\xi=-0.11$ this crossing occurs at $s=-0.22$ or $z=0.25$.}}
\end{figure}
Our results maybe compared with the minimal case that has been
investigated by Zhang and Zhu [14]. Indeed, by considering an
ordinary scalar field (quintessence), they have obtained a crossing
of the $\omega=1$ line running from $\omega>-1$ to $\omega<-1$ only
in the negative branch of the DGP scenario. However, we see here
that the presence of non-minimal coupling leads to a crossing
behavior in both branches of DGP-inspired scenario. With a single
quintessence scalar field this crossing runs from phantom phase to
quintessence phase. Figure $6$ shows the deceleration parameter in
two branches of the model. It is observed from the figure that an
accelerated phase will be occurs in the sufficiently high redshift
in the future.
\begin{figure}[htp]
\begin{center}
\includegraphics{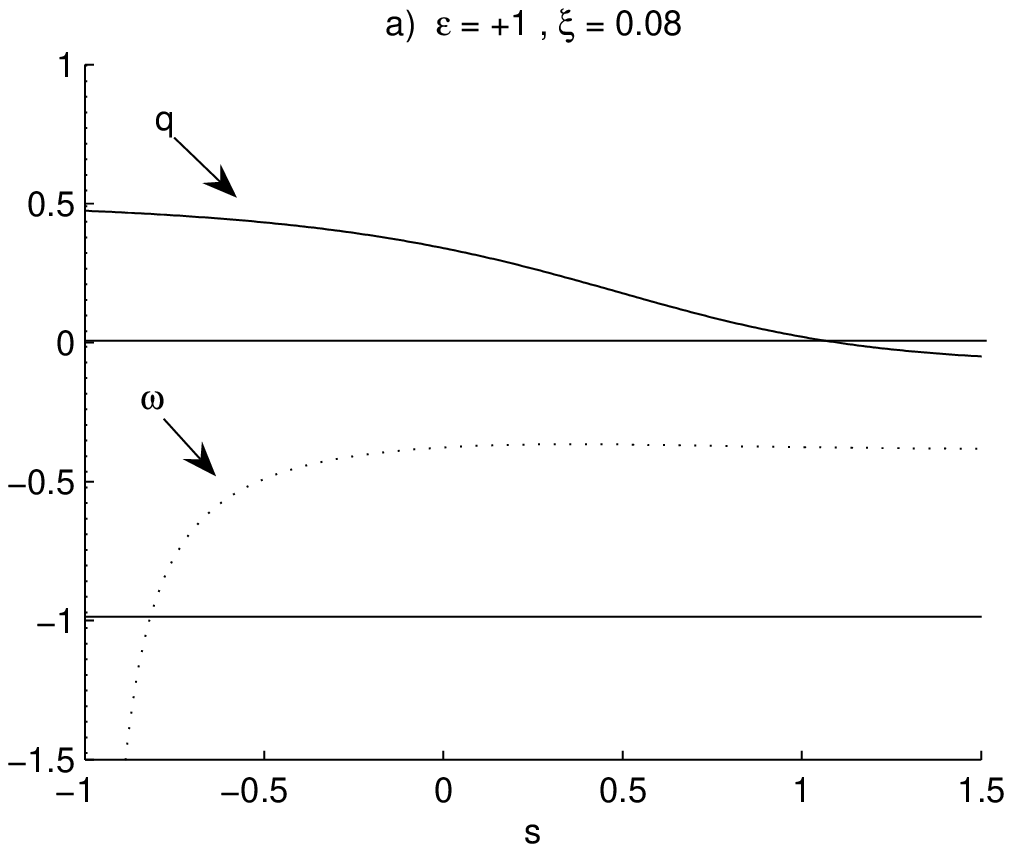} \vspace{5cm}\includegraphics{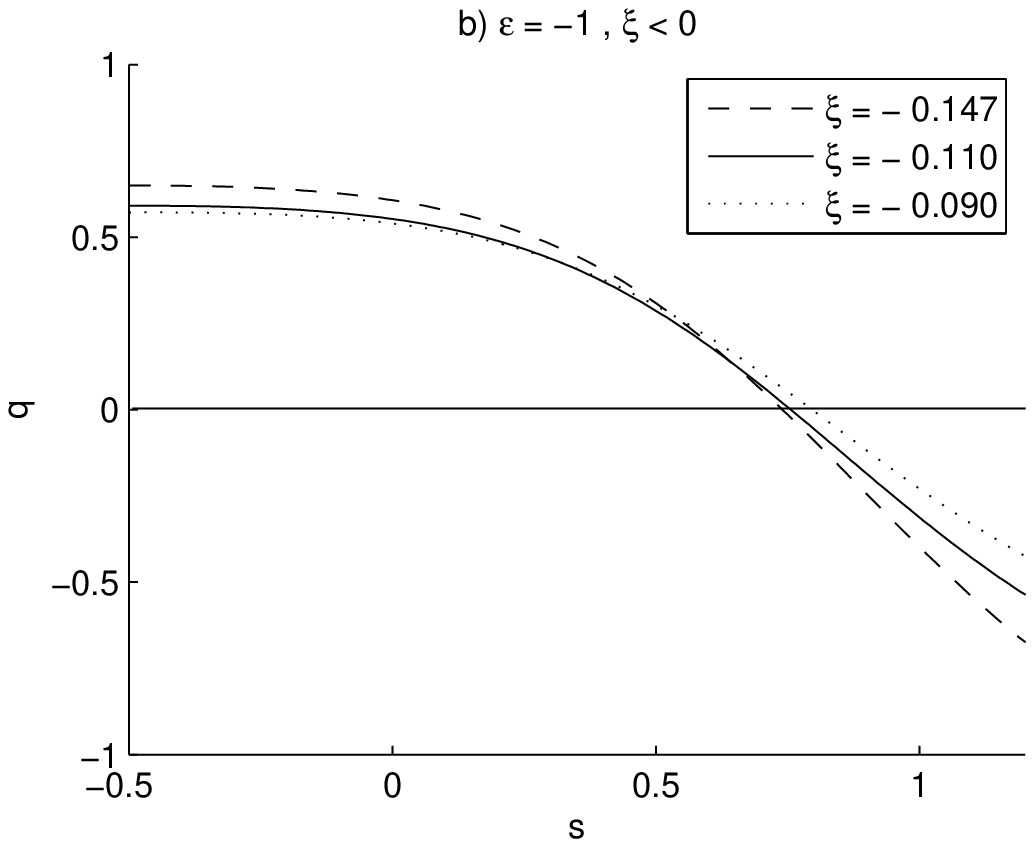}\vspace{5cm}
\end{center}
\vspace{2.5cm} \caption{\small {The deceleration parameter in
$DGP^{(+)}$ branch which vanishes for $\xi=0.08$ at $s\approx-0.82$
or $z\approx1.27$. Crossing of phantom divide line in this case  has
been shown with dotted curve. b) The deceleration parameter in
$DGP^{(-)}$ branch which vanishes at $s\approx 0.75$ for
$\xi=-0.110$.}}
\end{figure}
The role played by the parameter $\eta$ which is related to warped
effect has been shown in figure $7$. We should emphasize that from
this figure we see that for small values of $\eta$, the equation of
state parameter, $\omega$, crosses the cosmological constant line in
relatively small values of redshifts.
\begin{figure}[htp]
\begin{center}
\includegraphics{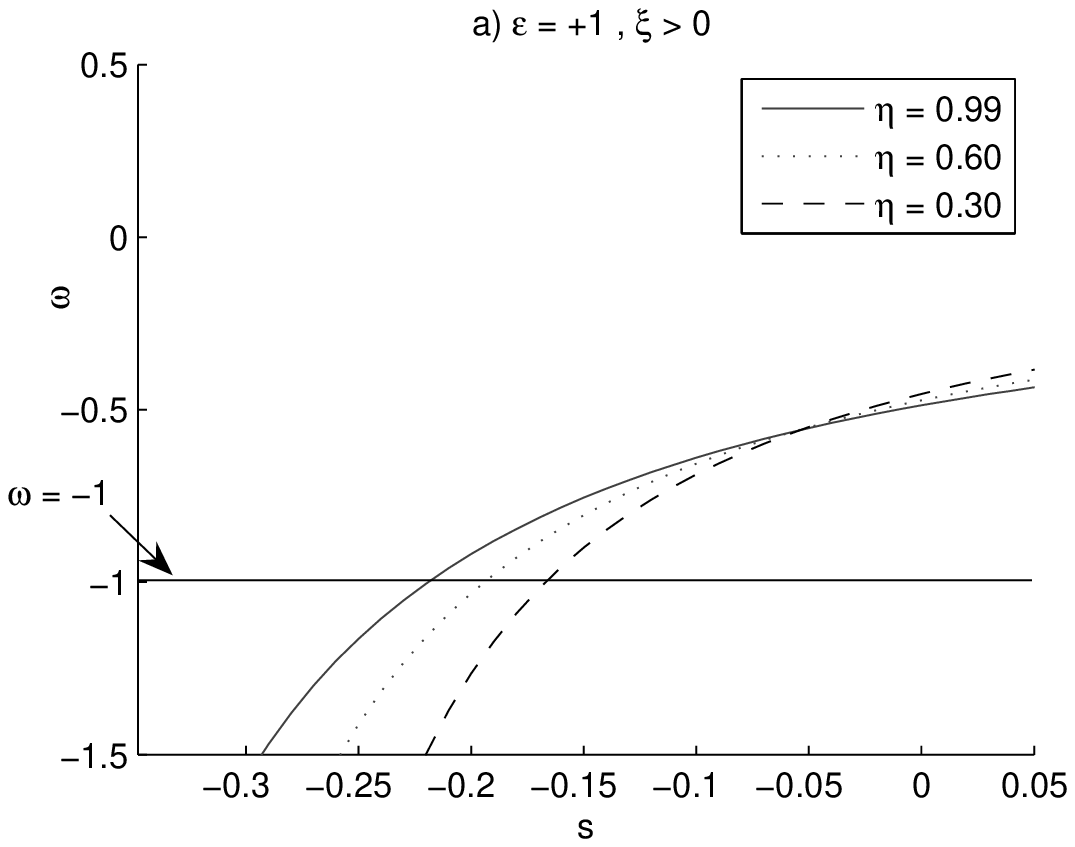} \vspace{5cm}\includegraphics{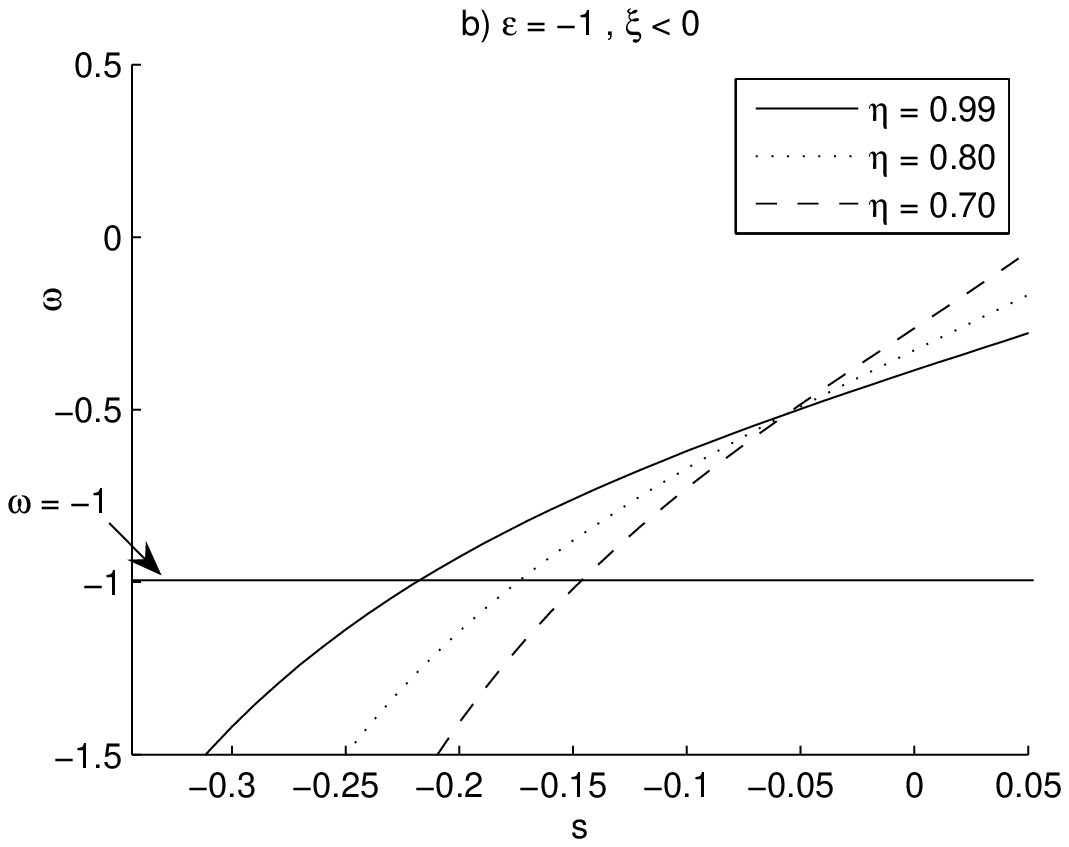}\vspace{5cm}
\end{center}
\vspace{2.5cm}
 \caption{\small {The role played by the warp effect ( via parameter $\eta$ ) on the crossing of the
 phantom divide line. For sufficiently small values of $\eta$, the equation of state
parameter, $\omega$, crosses the phantom divide line in relatively
small values of redshifts. a) In positive branch of the model with
$\xi=0.298$, the EoS parameter of dark energy crosses $\omega=-1$
line for $\eta=0.99$ at $s\approx-0.22 $ or $z\approx0.25 $. For
$\eta=0.6$ this crossing occurs at $s\approx-0.194$ or
$z\approx0.21$ and for $\eta=0.3$ occurs at $s\approx-0.167 $ or
$z\approx0.18$. b)In negative branch of the model with $\xi= -0.11$,
the EoS of dark energy crosses $\omega=-1$ line for $\eta=0.99$ at
$s\approx-0.22 $ or $z\approx0.25$. Also, for $\eta=0.8$ this
crossing occurs at $s\approx-0.174$ or $z\approx0.19$  and so on.}}
\end{figure}

In figure $8$, these results are shown in a three dimensional plot
of EoS parameter $\omega$ versus $s$ and $\eta$.
\begin{figure}[htp]
\begin{center}
\includegraphics{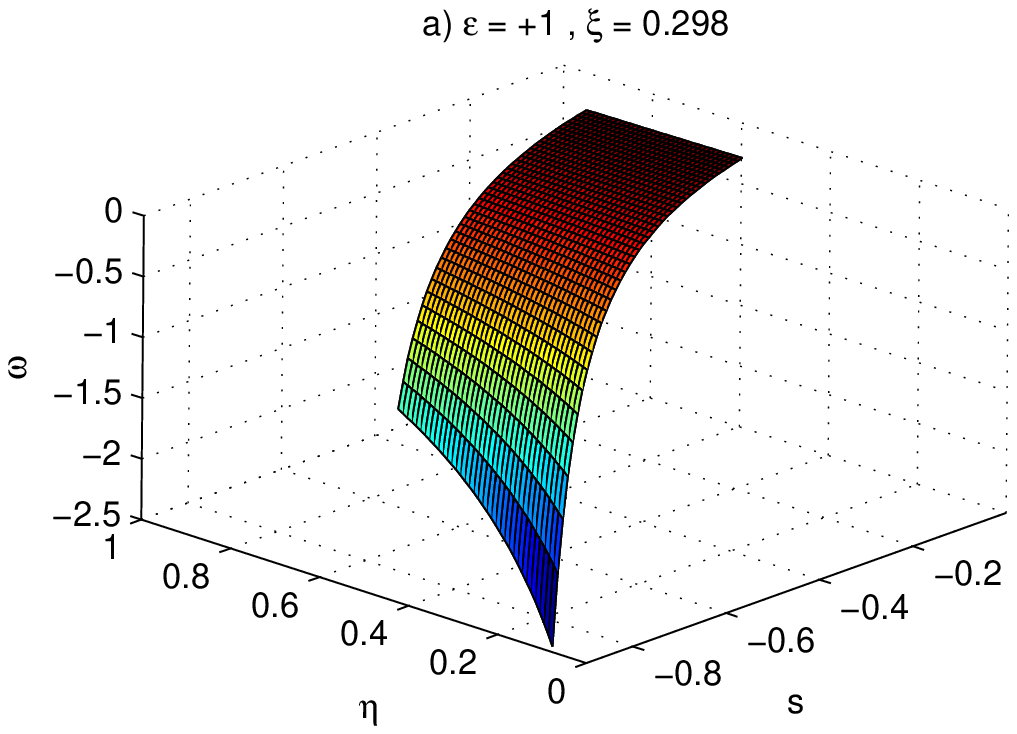} \vspace{5cm}\includegraphics{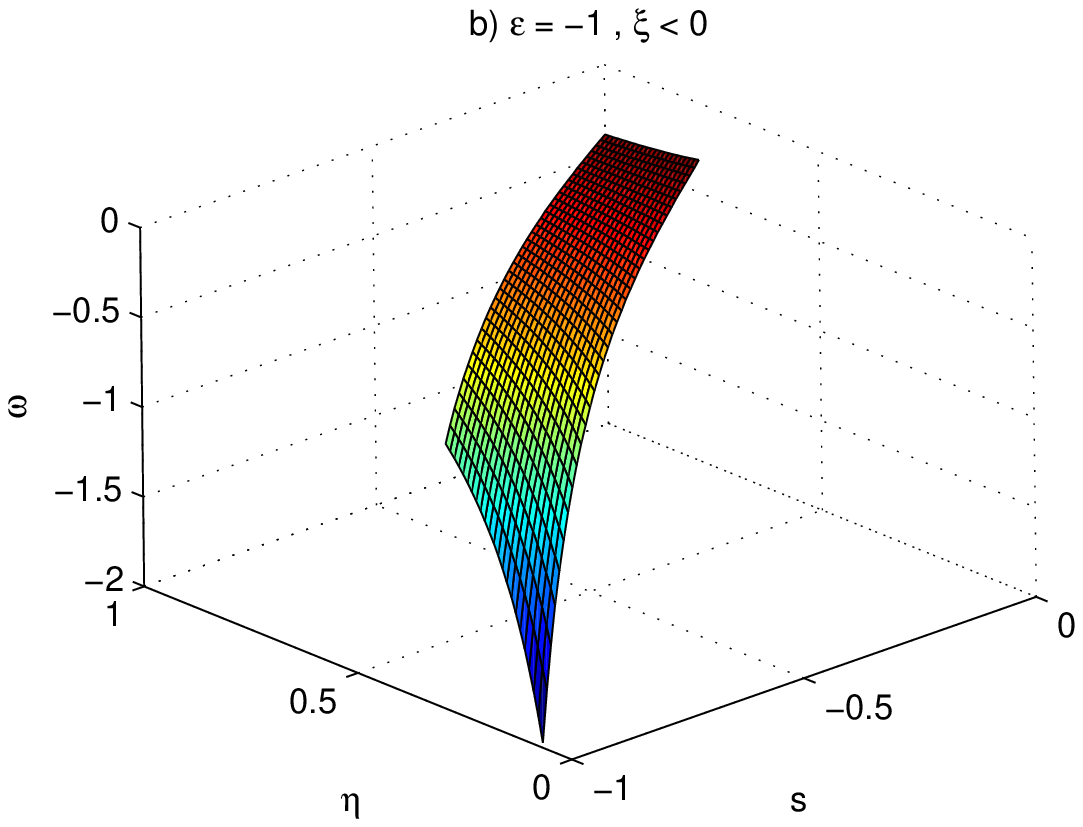}\vspace{5cm}
\end{center}
\vspace{2.5cm}
 \caption{\small {A three dimensional realization of the role played by the
 warp effect on the crossing of the phantom divide line. a) In
 $DGP^{+}$ branch of the model with $\xi=0.298$ b) In $DGP^{-}$
 branch with negative non-minimal coupling, $\xi=-0.11$. }}
\end{figure}

\newpage
\subsection{Phantom Field}
Now we investigate dynamics of equation of state with a phantom
field non-minimally coupled to induced gravity on the warped DGP
brane. The action of the model is given by
\begin{equation}
{\cal{S}}_{\sigma}=\int_{brane}d^{4}x\sqrt{-g}\Big[\frac{1}{2}\xi
R\sigma^{2}+\frac{1}{2}\partial_{\mu}\sigma\partial^{\mu}\sigma-V_2(\sigma)\Big].
\end{equation}
Energy density and pressure of this phantom field are defined as
\begin{equation}
\rho_{\sigma}=-\frac{1}{2}\dot{\sigma}^{2}+V_2(\sigma)-6\xi
H\sigma\dot{\sigma}-3\xi H^{2}\sigma^{2},
\end{equation}
and
\begin{equation}
p_{\sigma}=-\frac{1}{2}\dot{\sigma}^{2}-V_2(\sigma)+2\xi(\sigma\ddot{\sigma}+2\sigma
H\dot{\sigma}+\dot{\sigma}^{2})+\xi\sigma^2(2\dot{H}+3H^2)
\end{equation}
respectively. The equation of state parameter of this phantom field
is given by
$$\omega=-1+\frac{1}{\rho_{de}}\Bigg[-\dot{\sigma}^2+2\xi\Big(-H\sigma\dot{\sigma}
+\dot{H}\sigma^2+\sigma\ddot{\sigma}+\dot{\sigma}^2\Big)
+\Big[\dot{\sigma}^2+2\xi\Big(-H\sigma\dot{\sigma}
+\dot{H}\sigma^2+\sigma\ddot{\sigma}+\dot{\sigma}^2\Big)+\rho_{dm}\Big]$$
\begin{equation}
\Big[\varepsilon\eta\Big(A_{0}^2+2\eta\frac{\frac{1}{2}\dot{\sigma}^{2}+V_{2}(\sigma)-6\xi
H\sigma\dot{\sigma}-3\xi
H^{2}\sigma^{2}+\rho_{dm}}{\rho_0}\Big)^{-\frac{1}{2}}\Big]\Bigg]
\end{equation}
where
\begin{equation}
\ddot{\sigma}=\frac{6\xi\sigma\big[-\dot{\sigma}^2+2\xi\big(-H
\sigma\dot{\sigma}+\dot{\sigma}^2\big)+\rho_{dm}\big]\Big(\frac{b}{-2\big(\mu^{2}+\xi\sigma^2b\big)}\Big)+3H(\dot\sigma+4\xi\sigma
H)+\frac{\lambda_{2}}{\mu}V_2(\sigma)} {-1+\frac{6\xi^2\sigma^2
b}{\mu^2+\xi\sigma^2 b}}.
\end{equation}
Table $3$ shows the acceptable range of non-minimal coupling $\xi$
to have crossing of cosmological line in this case.
\newpage
\begin{table}
\begin{center}
\caption{Acceptable range of non-minimal coupling constraint by the
age of the universe and a positive root of $H$ (as given by equation
(26) with just one phantom field ) in order to have crossing of the
phantom divide line. } \vspace{0.5 cm}
\begin{tabular}{|c|c|c|c|c|c|c|c|}
  \hline
  \hline $\varepsilon$&$\xi$ &Acceptable range of $\xi$ & The value of $\xi$ for z=0.25  \\
  \hline +1&positive & $0.16\leq\xi\leq0.126$ &---  \\
  \hline +1&negative & $-0.106\leq\xi\leq-0.056$ & -0.091 \\
 \hline -1&positive &no crossing &--- \\
  \hline-1&negative &no crossing&---  \\
 \hline
\end{tabular}
\end{center}
\end{table}
Figure $9$ shows that in the positive branch of this DGP-inspired
scenario, the EoS of dark energy crosses the phantom divide line for
all values of non-minimal coupling parameter. It is worth noticing
that there is a very different behavior of such a crossing relative
to existing literature ( for instance [14]) . While the EoS of dark
energy crosses from below of cosmological line to its above for
positive values of $\xi$, with negative values of $\xi$ this
phenomenon occurs from above cosmological constant line to its
below. The figure also indicates that there is no crossing in the
negative branch of this setup for any range of non-minimal coupling
parameter. This result could be compared with the minimal case
obtained by Zhang and Zhu [14]. In fact in their framework, for
phantom field the EoS of dark energy crosses from $\omega<-1$ to
$\omega>-1$ in positive branch of DGP scenario. Here the situation
differs due to non-minimal coupling and warp effect.
\begin{figure}[htp]
\begin{center}
\includegraphics{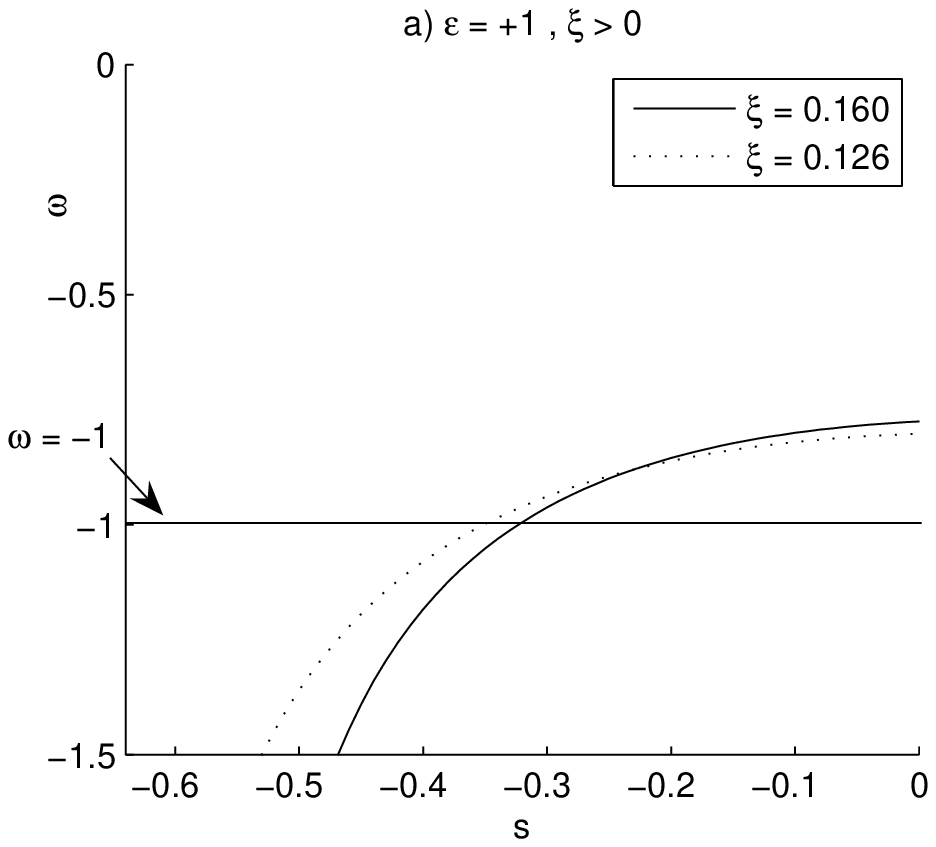} \vspace{5cm}\includegraphics{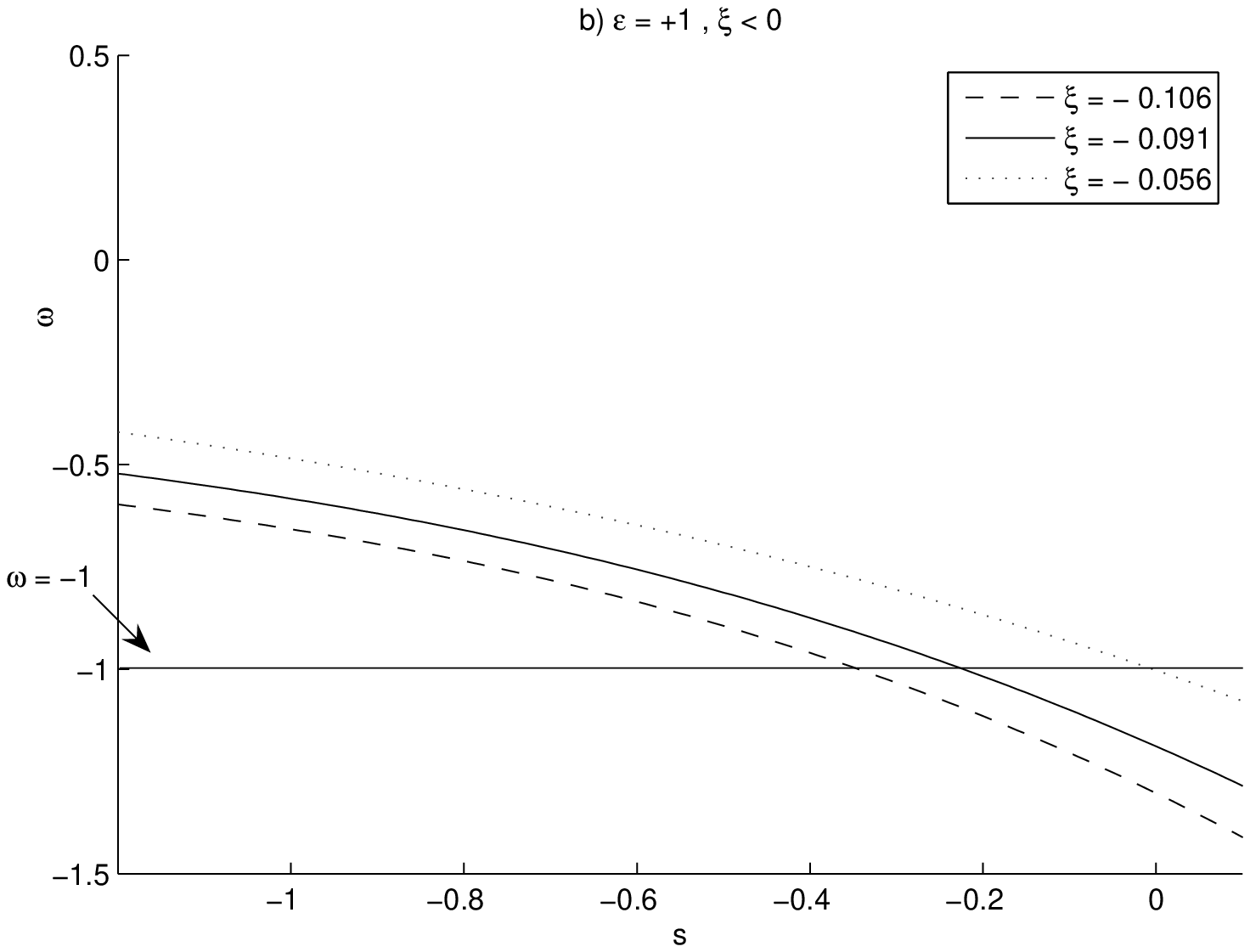}\vspace{5cm}\includegraphics{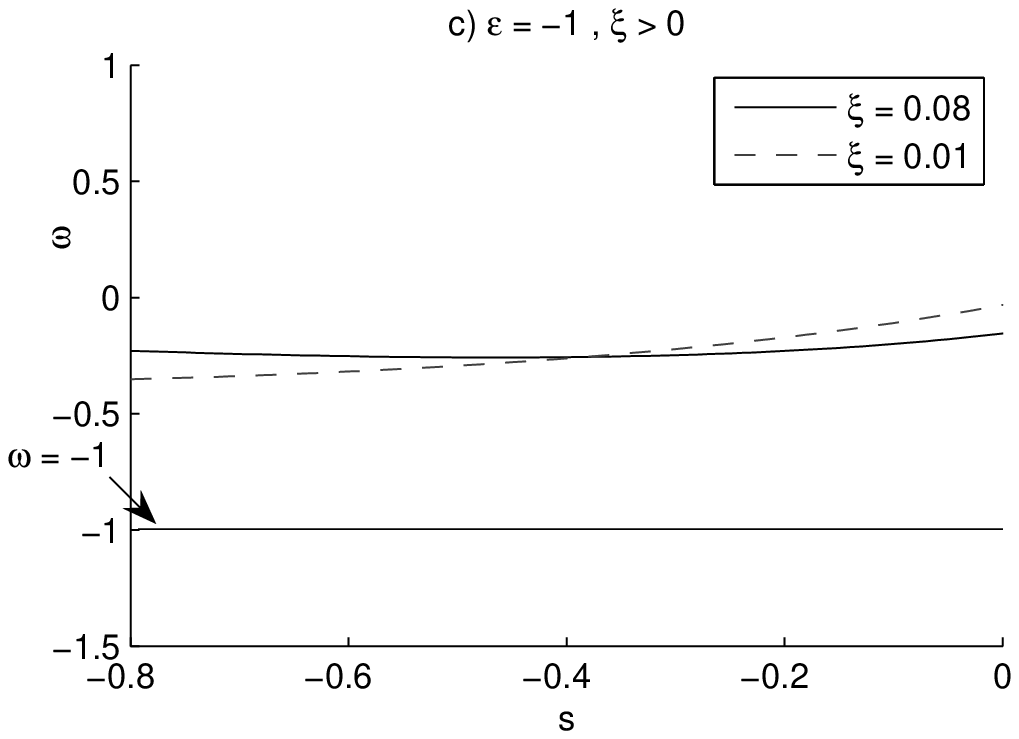}
\vspace{5cm}\includegraphics{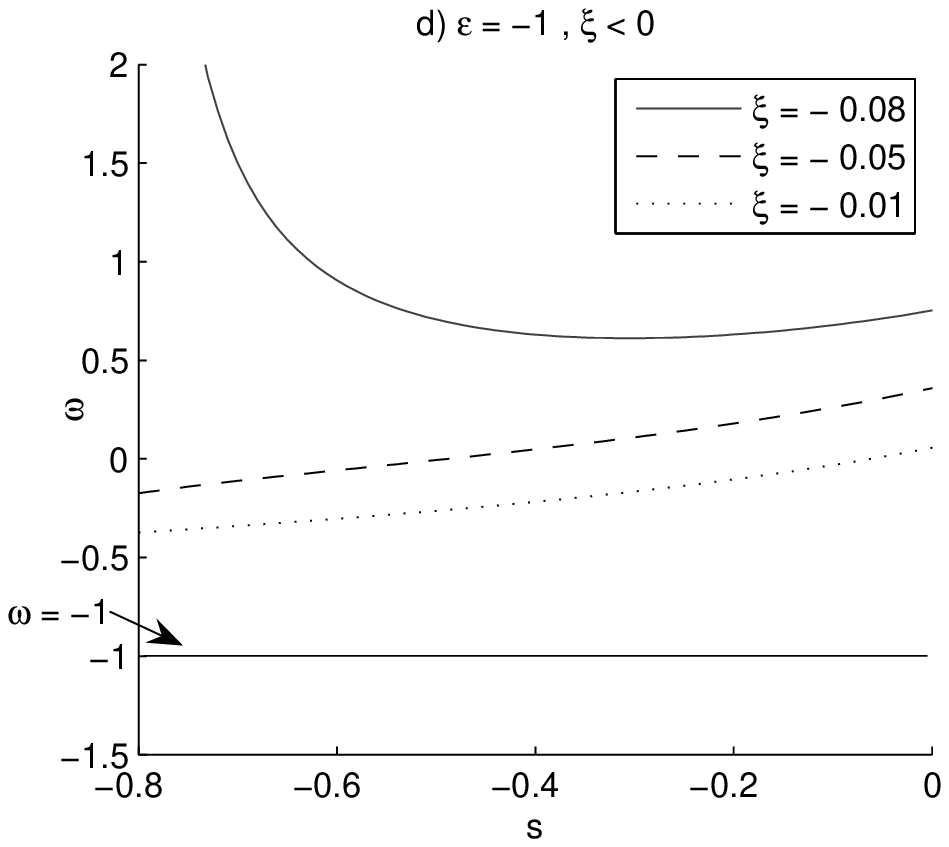}\vspace{0.5cm}
\end{center}
 \caption{\small {a)With positive values of nonminimal coupling,
 the EoS parameter of dark energy on the positive branch of the model crosses  the $\omega=-1$
 line. This crossing runs from phantom to quintessence phase.
 b) With negative values of nonminimal coupling,
 the EoS parameter of dark energy crosses the $\omega=-1$ line in positive branch of the model from
 quintessence to phantom phase.
 This crossing occurs for $\xi=-0.091$ at $s=-0.22$ or $z=0.25$.
 c)There is no crossing of phantom divide line in negative branch of the model with positive values of
 non-minimal coupling.  d) There is no crossing of phantom divide line in negative branch with
 negative values of non-minimal coupling.}}
\end{figure}

In figure $10$ we show the dynamics of deceleration parameter with a
non-minimally coupled phantom field on the warped DGP brane. The
result confirm that this parameter vanishes in sufficiently late
times of the universe evolution. We also plotted the
three-dimensional profile of EoS parameter $\omega$ versus $s$ and
$\eta$.

\begin{figure}[htp]
\begin{center}
\includegraphics{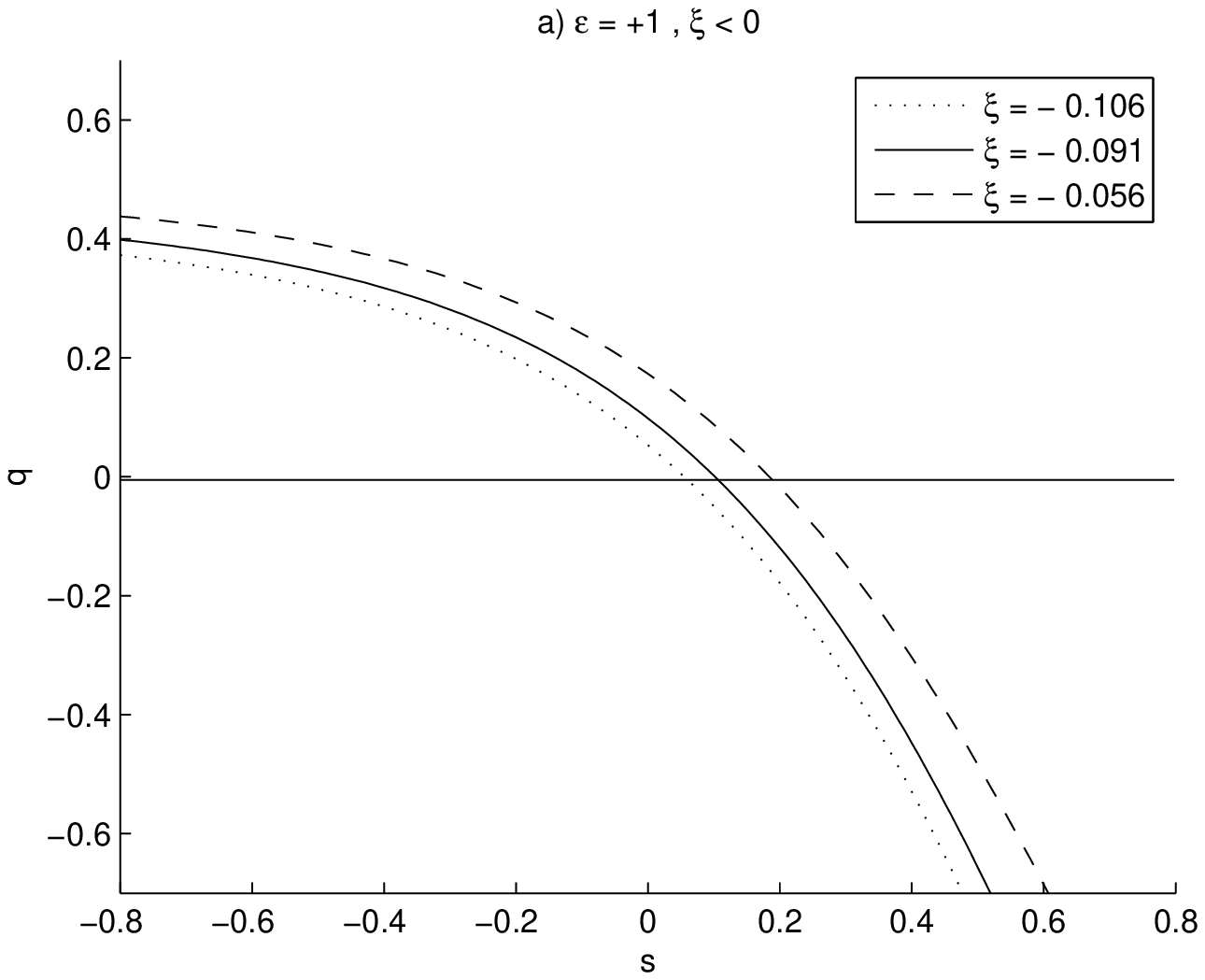} \vspace{5cm}\includegraphics{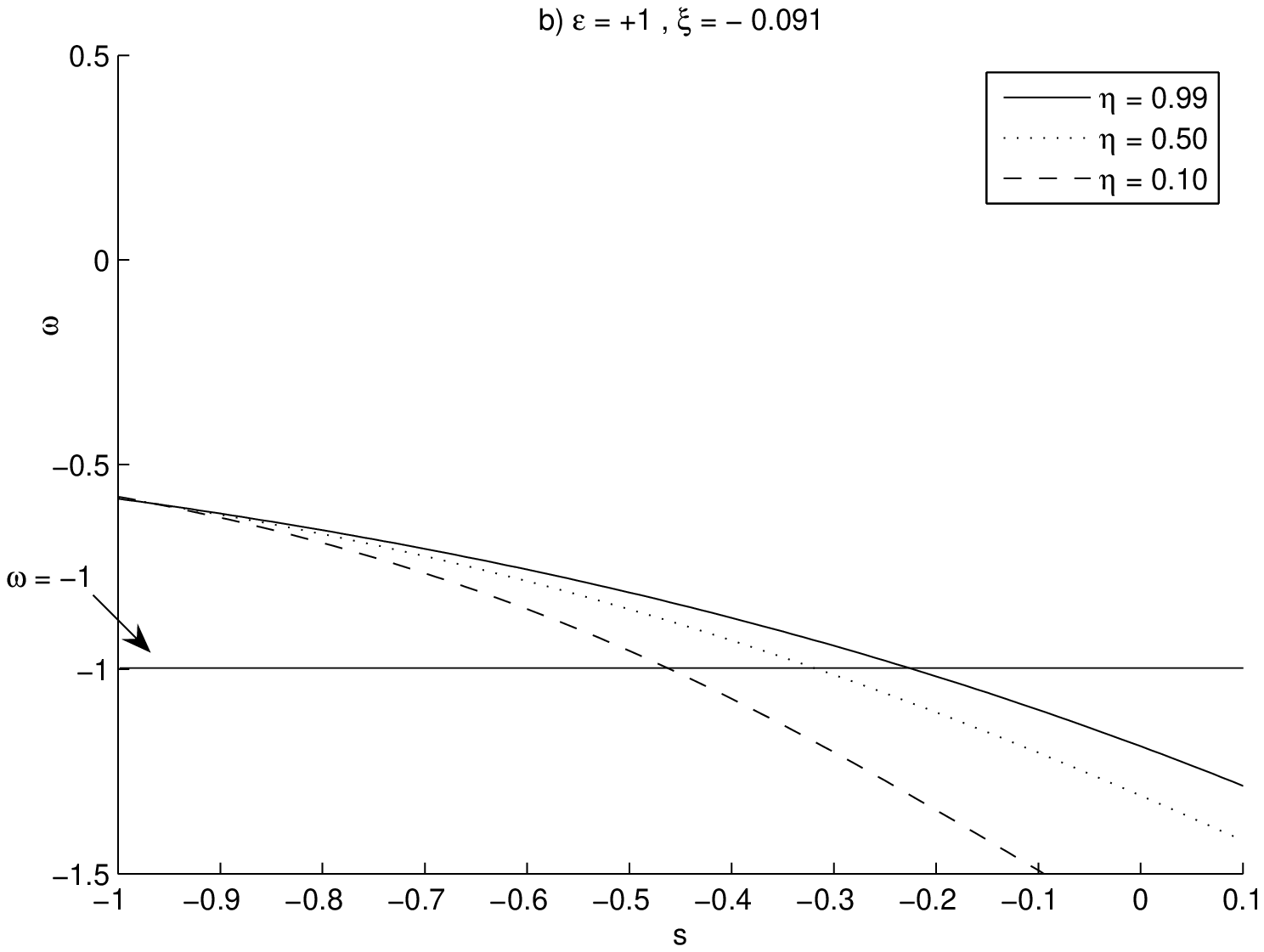}\vspace{5cm}\includegraphics{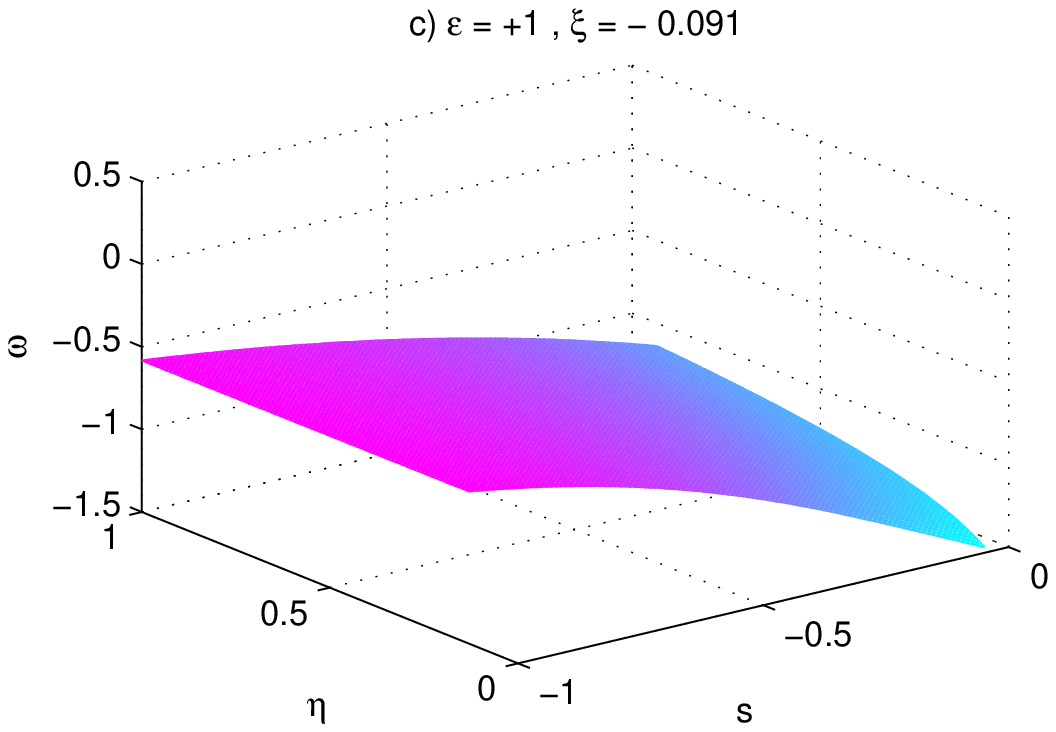}
\end{center}
\vspace{2.5cm}
 \caption{\small {a)Dynamics of deceleration parameter for the case with one phantom field.
 b)The role played by the parameter $\eta$ ( related to warp effect) on the crossing of the phantom divide line.
For sufficiently small values of $\eta$, equation of state
parameter, $\omega$, crosses the phantom divide line in relatively
large values of redshifts. In $DGP^{+}$ branch of the model and with
$\xi=-0.091$, the EoS of dark energy crosses the $\omega=-1$ line
for $\eta=0.99$ at $s\approx-0.22 $ or $z\approx0.25 $. For
$\eta=0.5$, this crossing occurs at $s\approx-0.315$ or
$z\approx0.37$ while for $\eta=0.1$ this occurs at $s\approx-0.46 $
or $z\approx0.58 $. c) Dynamics of equation of state parameter
versus $s$ and $\eta$. This figure is plotted for $DGP^{+}$ branch
of the model with $\xi=-0.091$. }}
\end{figure}

\section{Summary and Conclusion}
Light-curves analysis of several hundreds type Ia supernovae, WMAP
observations of the cosmic microwave background radiation and other
CMB-based experiments have shown that our universe is currently in a
period of accelerated expansion. An alternative to explain this
accelerated expansion is a multi-component dark energy with at least
one non-canonical phantom field. The analysis of the properties of
dark energy from recent observations mildly favor models where
$\omega=\frac{p}{\rho}$ crosses the phantom divide line, $\omega=-1$
in the near past. In this respect, construction of theoretical
frameworks with potential to describe positively accelerated
expansion and crossing of phantom divide line by equation of state
parameter is an interesting challenge. In this paper, we have
constructed a quintom dark energy model on a warped DGP brane. We
have investigated the crossing of the phantom divide line in this
setup in three cases: first we have considered a combined scenario
consist of two scalar fields as a realization of quintom model. This
model is built on a unified treatment of an ordinary (quintessence)
scalar field and a phantom field non-minimally coupled to the
induced Ricci scalar on the warped DGP brane. In this case we have
realized that cosmological constant line crossing occurs in both
branches of this DGP-inspired scenario with suitable values and
signs of non-minimal coupling parameter,($\xi$). For positive values
of $\xi$, the EoS parameter of dark energy crosses the phantom
divide line in $DGP^{+}$ branch of the model from below of
cosmological constant line (phantom phase) to its above
(quintessence phase). But, for negative values of $\xi$, this
crossing occurs in $DGP^{-}$ branch from above cosmological constant
line (quintessence phase) to its below (phantom phase). Secondly, we
have considered an ordinary scalar field (quintessence)
non-minimally coupled to the induced gravity. In this case we have
obtained crossing of the cosmological constant line in both
DGP$^{(\pm)}$ branches of the model. The crossing occurs for
positive branch by positive ranges of $\xi$ and for negative branch
by negative values of $\xi$.  We have compared our results with the
results of a similar analysis with minimal scalar field on the
ordinary DGP setup investigated by Zhang and Zhu [14]. They have
considered an ordinary scalar field (quintessence) on the DGP brane
and obtained a crossing of the phantom divide line from $\omega>-1$
to $\omega<-1$ only in negative branch of DGP scenario.
Consequently, by our analysis it can be concluded that the presence
of non-minimal coupling of scalar field and induced Ricci scalar on
the warped DGP brane leads to a crossing behavior in both branches
of this DGP-inspired scenario. And finally we have considered a
phantom scalar field non-minimally coupled to the induced Ricci
scalar on the brane. In this case the EoS parameter of dark energy
crosses the phantom divide line only in positive branch of model and
its behavior is sensitive to the sign of the non-minimal coupling
parameter. Indeed, the EoS of dark energy crosses cosmological
constant line from phantom phase to quintessence phase for positive
$\xi$; with negative $\xi$ this phenomenon occurs reversely (from
quintessence to phantom phase). By comparing our results with the
minimal case ( which for phantom field the EoS of dark energy
crosses from $\omega<-1$ to $\omega>-1$ in positive branch of DGP
scenario), we see that in our model this crossing behavior can occur
even in reverse direction depending on the sign of the non-minimal
coupling parameter.

\end{document}